\documentclass[aps,prd,twocolumn,groupedaddress,nofootinbib]{revtex4-2}
\usepackage[dvipsnames]{xcolor}
\usepackage[breaklinks,colorlinks,citecolor=blue,linkcolor=Green,urlcolor=blue]{hyperref}
\usepackage{gensymb}
\usepackage{amssymb}
\usepackage{amsmath}
\usepackage{float}
\usepackage{graphicx}
\usepackage{makecell}

\begin{document}


\title{Nature of M31 gamma-ray halo in relation to dark matter annihilation}


\author{Andrei E. Egorov}
\email[E-mail: ]{aegorov@runbox.com}
\affiliation{Nuclear Physics and Astrophysics Division, Lebedev Physical Institute, Leninskii prospect -- 53, 119333, Moscow, Russia}

\date{\today}

\begin{abstract}
The present work analyzes various aspects of M31 gamma-ray halo emission in its relation to annihilating dark matter (DM). The main aspect is the predicted effect of asymmetry of the intensity of emission due to inverse Compton scattering (ICS) of a possible population of relativistic electrons and positrons ($e^\pm$) in the galactic halo on starlight photons. This asymmetry is expected to exist around the major galactic axis, and arises due to anisotropy of the interstellar radiation field and the inclination of M31. ICS emission and its asymmetry were modeled by GALPROP code for the trial case of $e^\pm$ generated by annihilating weakly interacting massive particles (WIMPs) with various properties. The asymmetry was obtained to appear at photon energies above $\sim$ 0.1 MeV. Morphological and spectral properties of the asymmetry were studied in detail. Potential observational detection of the asymmetry may allow to infer the leptonic fraction in the emission generation mechanism, thus providing valuable inferences for understanding the nature of M31 gamma-ray halo emission. Specific asymmetry predictions were made for the recently claimed DM interpretation of the outer halo emission. The paper also studied the role of secondary -- ICS and bremsstrahlung -- emissions due to DM annihilation for that interpretation. And, finally, the latter was shown to be somewhat restricted by the recently derived WIMP constraints from radio data on M31.
\end{abstract}


\maketitle

\section{\label{sec:i}Introduction and motivation}

M31 (Andromeda galaxy) is the closest large spiral galaxy. Its proximity allows to study in detail a wide variety of astrophysical phenomena under a view and environment, which are alternative to our own Galaxy, Milky Way (MW). This paper concerns the results of gamma-ray band observations of M31 in their possible relation to annihilating dark matter (DM). M31 was detected in gamma rays for the first time by Fermi-LAT during the first years of its operation \cite{2010A&A...523L...2A}. Later, more observational data was accumulating, which enabled certain detailization of both the emission spectrum and morphology. Thus, \cite{2017ApJ...836..208A} reported that the gamma-ray emission source at M31 center is extended, has the angular radius $\approx 0.4\degree$ and presumably has a uniform brightness distribution. The emission spectrum was measured up to $\approx$ 10 GeV. Then \cite{2021PhRvD.103h3023A} (among other works on the subject) confirmed the cited results of \cite{2017ApJ...836..208A} and also reported a possible presence of another emission component, which extends up to $\approx 1\degree$. At about the same time, \cite{2019ApJ...880...95K} also found out a tentative presence of very large outer gamma-ray halo, which extends up to at least $\approx 9\degree$. And very recently, \cite{2023ApJ...945L..22X} elaborated that the central source, which was previously thought to be extended with the radius $0.4\degree$, in fact represents two point sources: one is in the center and one is $\approx 0.4\degree$ away from the center. And it is not very clear whether the non-central source belongs to M31.

Simultaneously with the observational progress briefly described above, extensive theoretical modeling of the possible gamma-ray emission mechanisms in M31 was developing. A wide variety of emission sources has been proposed: a population of millisecond pulsars (MSPs) \cite{2022MNRAS.516.4469Z}, cosmic rays (CRs) \cite{2019PhRvD.100b3014M,2021PhRvD.104l3016D} and DM annihilation (e.g., \cite{2021PhRvD.103b3027K}). It is very possible that more than one emission mechanisms work together, and different processes are responsible for the emission generation in different regions of the galaxy.

Many works studied the possibility of presence of DM contribution in the gamma-ray emission \cite{2016JCAP...12..028L,2018PhRvD..97j3021M,2019PhRvD..99l3027D,2021PhRvD.103b3027K,2022MNRAS.516.4469Z} and derived the respective constraints. One brief conclusion from these studies is  impossibility to explain all the emission by annihilating DM only. Thus, the fit of the inner halo (IH) region spectrum requires the mass of DM particle (traditional weakly interacting massive
particles (WIMPs) are being considered) to be very small -- $m_x \approx (6-11)$ GeV according to \cite{2018PhRvD..97j3021M}, while the outer halo (OH) region spectrum can be fitted by heavier WIMP with $m_x \approx (45-72)$ GeV \cite{2021PhRvD.103b3027K}. But this is an absolutely natural situation: we would primarily expect for DM contribution to be minor, while the majority of emission is generated by usual astrophysical processes, like it is in our own Galaxy. And it is very interesting and promising to understand the emission nature in detail, which may eventually lead to a robust detection of DM signal among other emission components. Thus, M31 studies in the gamma-ray band represent a valuable direction in the field of DM indirect searches.

One potential emission generation mechanism is leptonic, i.e. through the inverse Compton scattering (ICS) process between the photon field in the galaxy and relativistic electrons/positrons. This mechanism has a big relevance: \cite{2019PhRvD.100b3014M,2021PhRvD.104l3016D} showed that both the IH and OH emissions could be explained by CR interactions, when a significant contribution comes from ICS of CR $e^\pm$. The emission due to WIMP annihilation (or decay) would also necessarily have the leptonic (secondary) component; since besides the prompt gammas, the annihilation produces relativistic $e^\pm$ too (DM $e^\pm$ hereafter). However, among all the works cited above, only \cite{2016JCAP...12..028L} paid attention to the ICS emission component due to WIMPs.

The main goal of this work is to conduct comprehensive theoretical modeling of the interesting effect of ICS emission intensity asymmetry between M31 hemispheres. This effect was pointed out for the first time in \cite{2020arXiv201104689B} (see Fig. 2 there): we view M31 under the certain inclination angle; hence, there must be some difference in the upscattered photon spectra from the regions above and below M31 major axis, since the starlight radiation field in any disk galaxy has a significant anisotropy. The physical nature of such asymmetry will be explained in more details below. The authors of \cite{2020arXiv201104689B} provided just simple analytical estimates of this effect, concluding that an average upscattered photon energy would differ by $\sim$ 10\% between the hemispheres. Here I aim to study this asymmetry effect much more precisely and in detail, since the former may potentially provide a valuable observational test for the physical nature of the gamma-ray emission from the halo. GALPROP code \cite{GP} was employed for the modeling of ICS emission in M31. As the source of $e^\pm$ population in the halo, WIMP annihilation was assumed in connection with the results of \cite{2021PhRvD.103b3027K}, where WIMP annihilation in both MW and M31 was shown to be the plausible explanation of the gamma-ray emission toward the OH. The study here is majorly theoretical and does not aim so far to derive any observational constraints or implications, although some minor qualitative discussion of the observational prospects is given. Sec. \ref{sec:as} below is dedicated to the asymmetry effect and represents the main content of this paper. Then Secs. \ref{sec:sec} and \ref{sec:dm} discuss somewhat different but important aspects of DM interpretation of the OH emission. The authors of \cite{2021PhRvD.103b3027K} performed the fit of the OH by only the prompt gamma-ray emission from annihilating WIMPs and did not take into account the secondary (ICS and bremsstrahlung) emissions. Sec. \ref{sec:sec} analyzes the role of secondary emission contributions due to WIMPs. Sec. \ref{sec:dm} debates the relation between the OH fit and WIMP radio constraints. And Sec. \ref{sec:con} summarizes the findings.

In general, this work continues our series of papers \cite{2013PhRvD..88b3504E,2022PhRvD.106b3023E,2022arXiv220814186E} dedicated to WIMP searches/constraints in M31 and utilizes the methodology developed previously there. M31 model parameters assumed here are the same as in \cite{2022PhRvD.106b3023E}, if it is not explicitly stated otherwise.  

\section{\label{sec:as}ICS emission asymmetry}

\begin{figure}[h]
	\includegraphics[width=1.05\linewidth]{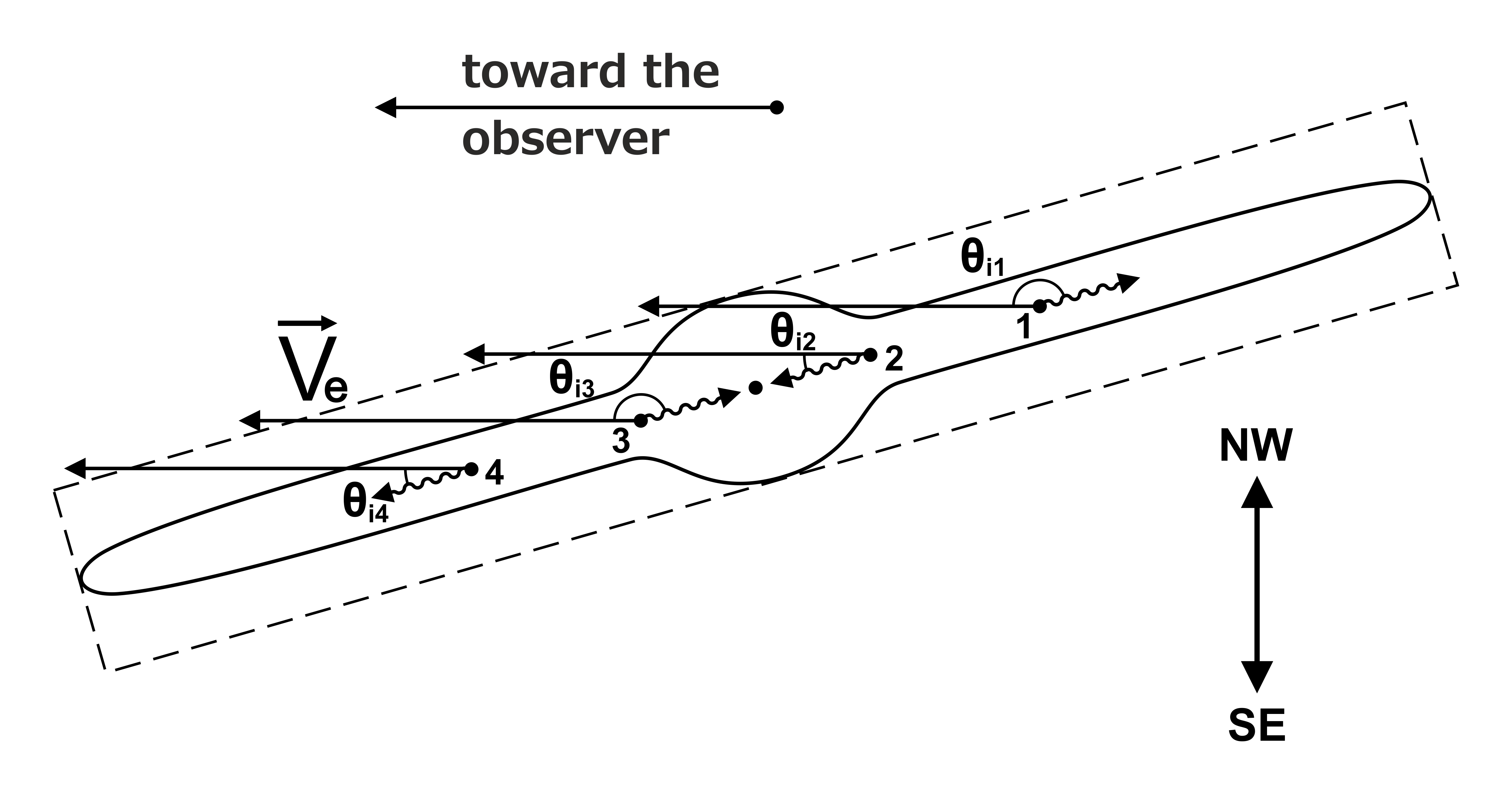}
	\caption{\label{fig:ics} The schematic central vertical section of M31 with four trial points of interest in regard to ICS process. The dashed line denotes the approximate boundary of the assumed diffusion zone; $\vec{V_e}$ -- the velocities of $e^\pm$, which produce the upscattered photons visible for the observer; and the wavy lines -- the purportive slightly predominant directions of ISRF photons at the corresponding points. }
\end{figure}

An interstellar radiation field (ISRF), i.e. the photon field produced by stars, gas and dust, is anisotropic at some level in any disk galaxy due to its non-spherical shape. Let us imagine that a disk galaxy, particularly M31 in our case, possesses a population of relativistic $e^\pm$ of any origin, they produce gamma rays through ICS at ISRF, and we observe the galaxy in the gamma-ray band from aside under the certain inclination angle -- see Fig. \ref{fig:ics}. In our case $E_e > E_\gamma \gg E_{\gamma 0}$, where $E_e$ is the electron or positron energy before scattering, $E_{\gamma 0}$ and $E_\gamma$ are the photon energies before and after scattering. Such energy ratio implies, that the momentum of upscattered photon would have almost exactly the same direction as the initial $e^\pm$ momentum. This fact is quite obvious from just a general intuition, but can also be obtained from the equation for the differential cross section of the ICS process derived in quantum electrodynamics (e.g., [\cite{LL4}, Eq. (86.6)]). Hence, we would see the gamma-ray photons produced by $e^\pm$, which move toward the observer. Then we can deduce from Fig. \ref{fig:ics}, that the effective average angle between the initial momenta of photons and emitting $e^\pm$ slightly differs for the lines of sight lying above and below the major axis of the galaxy. The energy of upscattered photons strongly depends on that angle through the following relation (taken from [\cite{Ginzburg}, Eq. (16.54)]):
\begin{equation}\label{eq:Eg}
E_\gamma = \frac{E_{\gamma 0}(1-\beta_e \cos \theta_i)}{1-\beta_e \cos \theta_f + (1-\cos(\theta_i - \theta_f))E_{\gamma 0}/E_e},
\end{equation}
where $\beta_e \equiv V_e/c$, $\theta_i$ is the angle between the initial momenta of $e^\pm$ and photon, $\theta_f \approx 0$ is the angle after the collision (which nearly vanishes according to the considerations above). Therefore, we can expect slightly different ICS emission spectra from regions of interest (ROIs), which are located above and below the major axis, and are symmetric with respect to it. This is the ICS emission asymmetry effect in its essence. Roughly speaking, one hemisphere is expected to be brighter than the other at the same photon energy. Such effect does not exist for our own Galaxy, since we are located at its plane and view the ISRF from both hemispheres symmetrically. However, for M31 the asymmetry may be substantial. And potential observational studies of this asymmetry may allow to derive or constrain the leptonic fraction in the total gamma-ray emission, facilitating the unraveling of the emission mechanism.

The asymmetry effect is modeled here specifically for the case of $e^\pm$ population produced by annihilating DM, whose properties fit the outer gamma-ray halo of M31 according to \cite{2021PhRvD.103b3027K}. The primary motivation for such choice is to build precise and testable predictions for this specific model of the OH emission. At the same time, we may expect rather similar asymmetry properties for other cases of $e^\pm$ population, since the properties of the latter primarily influence the ICS emission intensity amplitude and spectral shape rather than the asymmetry. The latter is primarily defined by the galaxy inclination and ISRF directionality, which are the same for any $e^\pm$ population. Hence, in the first approximation, we may extrapolate the obtained asymmetry picture from our particular case of interest to other cases. Nevertheless, I computed the asymmetry properties for some alternative $e^\pm$ populations too in order to study the parameter dependence to a certain extent. It is not easy to compute a large grid of various $e^\pm$ model populations due to the computational heaviness of the task. 

\subsection{\label{sec:mod}Modeling the ICS emission due to WIMP annihilation}
This subsection describes the details of the ICS emission modeling procedure. WIMP mass was set to $m_x = 60$ GeV as the main value of interest. This is the approximate average value among those derived in \cite{2021PhRvD.103b3027K} for the fit of OH. The primary annihilation products are assumed to be $b\bar{b}$ quark pairs. The annihilation cross section is set to the thermal value $\langle \sigma v \rangle (m_x = 60~\text{GeV}) = 2.1 \times 10^{-26}~\text{cm}^3$/s according to \cite{sv}. Actually, \cite{2021PhRvD.103b3027K} found very large uncertainty range for the cross section value, which fits the OH (Table II there): $\langle \sigma v \rangle \sim (10^{-26}-10^{-23})~\text{cm}^3$/s depending on a variety of their model parameters and assumptions. Only the thermal cross section is mainly considered here, because it represents the most natural and motivated value in cosmology and particle physics. But also the intensity of emission of any kind due to WIMP annihilation is linearly proportional to the cross section; hence, the intensity can be easily rescaled for an arbitrary cross section value. And the asymmetry, defined as the intensity ratio between the hemispheres, would not depend on the cross section at all.

Regarding DM density profile, \cite{2021PhRvD.103b3027K} utilized two profiles for the fit: Einasto profile from \cite{2012A&A...546A...4T} and Navarro-Frenck-White (NFW) profile from \cite{2019ApJ...880...95K,2021PhRvD.103b3027K}. These profiles yield rather similar densities (see [\cite{2021PhRvD.103b3027K}, Fig. 3]). I utilized Einasto profile for main calculations, though some alternative profile with lower density was substituted too (see Table \ref{tab:as} below) in order to study the parameter dependence. Meanwhile, Einasto profile exactly corresponds to MAX DM density profile in \cite{2022PhRvD.106b3023E} (its parameters can be seen in Table III there). MIN-MED-MAX here have the traditional meaning of parameter configurations, which provide respectively the minimal, medium and maximal emission intensities due to DM.    

Another important aspect is DM annihilation rate boost due to substructures in DM halo. This aspect is highly uncertain. The boost factor has the biggest influence on the annihilation rate at the largest radii $\sim$ 100 kpc. The authors of \cite{2021PhRvD.103b3027K} quantify the uncertainty in the boost factor by employing of four various models: the smooth density profile (i.e., no boost at all) and "low--mid--high" boost configurations (Table II there). A higher boost implies a lower annihilation cross section required for the outer gamma-ray halo fit. The boost factor radial profile adapted here is based on \cite{2010PhRvD..81d3532K} and closely resembles the "high" case in \cite{2021PhRvD.103b3027K}. In general, we do not expect a high influence of details of the boost factor model on the characteristics of the secondary emissions due to WIMPs, which we are interested in. This is because DM $e^\pm$ live only in the diffusion zone, which is typically assumed to extend up to at most $R \approx 20$ kpc (e.g., \cite{2012JCAP...01..005F}). And the annihilation rate boost factor is not yet significant at such distances (see, e.g., [\cite{2010PhRvD..81d3532K}, Fig. 4]). Thus, although the boost factor is included in the calculations here, it does not play a big role for the final results. But speaking technically, the model setup here was made to be consistent with \cite{2021PhRvD.103b3027K} in the following sense: the thermal annihilation cross section requires the "high" boost at least in MW for the OH fit according to Table II there; MW and M31 have quite similar DM halos; hence, the high substructure boost in MW likely implies the same for M31.

The computation of all the emission maps was performed by GALPROP code v57 \cite{2022ApJS..262...30P} paired with the addition \cite{github} developed by the author. This addition enables a precise computation of all DM-related particles and emissions, and incorporates the energy spectra of stable annihilation products (at injection) from PPPC 4 DM ID resource \cite{PPPC,2011JCAP...03..051C,2011JCAP...03..019C}. GALPROP solves the transport equation for DM $e^\pm$, computes their emissivity and integrates it along a line of sight, yielding emission intensity sky maps. GALPROP provides the powerful capability to compute the ICS emission precisely in the frame of full anisotropic formalism, which is described in \cite{2000ApJ...528..357M}. Indeed, GALPROP had to be specifically adapted for modeling of M31. In this aspect, the experience, which was developed in the frame of work \cite{2022PhRvD.106b3023E}, was widely used. Our model is 2D, i.e. the axial symmetry of the galaxy is assumed. An important model element for the ICS emission computation is ISRF, i.e. the model of distribution of the intensity of field target photons over energy, spatial coordinates and direction. In general, GALPROP naturally includes three distinct components of ISRF: CMB, a far-infrared (IR) emission from dust and an optical emission from stars. The standard MW 2D ISRF model created by GALPROP authors \cite{2005ICRC....4...77P} was utilized for M31 here with one modification: the energy densities of IR and optical components were globally rescaled according to the ratio of M31 and MW IR/optical total luminosities. GALPROP lacks the dedicated model of M31 ISRF. But MW and M31 are rather similar galaxies in their structure and size. Hence, it is a reasonable approximation to extrapolate MW ISRF to M31 for estimation purposes. And the ratio of M31 and MW luminosities serves as an estimator of the ratio of the photon field energy densities (again assuming the same size for both galaxies), since a radiation energy density is linearly proportional to an intensity. Table \ref{tab:lgm} provides the relevant luminosities of both galaxies. The corresponding ISRF energy density global rescaling factors are listed in Table \ref{tab:gp} together with other important GALPROP model parameters. It is implicitly assumed here that there are no causes for the ICS emission asymmetry other than the galaxy inclination. It means that $e^\pm$ population and ISRF are constructed in such a way, that they are both axisymmetric and symmetric with respect to the galactic plane. These are generally natural assumptions for a disk galaxy modeled in 2D. And all the used ISRF models were empirically checked to satisfy those symmetries by making test GALPROP runs with the observer placed at the galactic plane. These test runs yielded no ICS asymmetry, as anticipated.
\begin{table}[h]
	\caption{\label{tab:lgm}The total luminosities and gas masses of MW and M31 with respective references.}
	\begin{ruledtabular}
		\centering
		\begin{tabular}{ccc}
			Parameter & MW & M31 \\                     
			\hline
Optical (V-band)                 & & \\
luminosity [$L_\odot$]           & $2.1\times 10^{10}$ {\cite{2009A&A...505..497Y}} & $2.7\times 10^{10}$ {\cite{2009A&A...505..497Y}} \\
IR (dust) luminosity [$L_\odot$] & $7.4\times 10^{9}$ {\cite{1997ApJ...480..173S}} & $4.3\times 10^{9}$ {\cite{2014ApJ...780..172D}} \\
Gas mass [$M_\odot$]             & $7\times 10^{9}$ {\cite{2009A&A...505..497Y}} & $8\times 10^{9}$ {\cite{2009ApJ...695..937B,2006A&A...453..459N}} \\
		\end{tabular}
	\end{ruledtabular}
\end{table}		

Another essential model ingredient for the computation of ICS emission due to DM $e^\pm$ is their cooling and propagation. The cooling mechanisms include ICS, synchrotron, bremsstrahlung radiative energy losses; Coulomb scattering and ionization losses. The respective energy loss rates (as they are implemented in GALPROP) are written out in [\cite{2022PhRvD.106b3023E}, Eqs. (10)--(13)]. The ICS energy loss rate is defined by the energy density of ISRF, which is described above. The synchrotron loss rate is defined by the magnetic field (MF) strength. M31 MF model developed in \cite{2022PhRvD.106b3023E} (Sec. IIIC there) was naturally adopted here. Specifically, MED (i.e., medium) MF configuration with the central field strength of 50 $\mu$G was used as the base scenario here. The energy losses of other kinds require to set the gas distributions in M31. The former is described in [\cite{2022PhRvD.106b3023E}, Eq. (14), Appendix]. Regarding $e^\pm$ propagation (prop.) parameters, the experience gained in \cite{2022PhRvD.106b3023E} was again utilized: MED propagation parameter configuration, described in Sec. IIID there, was adopted as the base scenario. Thus, $e^\pm$ population is assumed to be residing and emitting inside the diffusion cylinder with the radius $r_\text{max}$ = 20 kpc and the half-height $z_\text{max}$ = 2.7 kpc. Indeed, such cylinder is a conventional modeling abstraction -- there is no sharp physical boundary at the cylinder edges. In reality, MF vanishes smoothly, and, hence, the diffusion coefficient diverges gradually too near the boundaries of the diffusion zone. For this reason, our model calculations of the ICS emission intensity might be unreliable for the lines of sight passing near the edges of the diffusion zone.

M31 was placed at the real distance in the model setup here, which differs from the model in \cite{2022PhRvD.106b3023E}, where a smaller than real distance was used in order to reduce computational efforts without a loss of precision. But it would be unreliable to alter the distance in the current task of the ICS emission asymmetry calculation, since such alteration may introduce unphysical asymmetry; geometrical proportions are more important here. The real distance required, in turn, very high angular resolution of sky maps, which were computed in the traditional HEALPix format \cite{2005ApJ...622..759G}. It was found empirically, that $\approx$ 1 arcmin resolution is sufficient for mitigating the finite pixelization effects on relevant angular scales. HEALPix resolution parameter $N_\text{side}$ was set to $2^{12}$ = 4096 in order to achieve the mentioned pixel size. The anisotropic ICS computation in GALPROP is heavy, and so high angular resolution required to employ the computing cluster with 64 CPU cores in order to compute the whole task in a few weeks.
\begin{table}[h]
	\caption{\label{tab:gp}Values of various parameters used in GALPROP.}
	\begin{ruledtabular}
		\centering
		\begin{tabular}{lc}
			Parameter & Value \\
			\hline
Radius of the diffusion cylinder $r_{\text{max}}$ & 20 kpc \\
Spatial grid step size $\Delta r = \Delta z$ & 0.05 kpc \\
$e^\pm$ propagation energy range & $(5\times 10^{-4} - 1)m_x c^2$ \\
Energy grid step increment & 1.1 \\
Optical ISRF energy density & \\
factor with respect to MW   & 1.3 \\
IR ISRF energy density & \\
factor with respect to MW & 0.57 \\
HEALPix resolution of maps $N_{\text{side}}$ & $2^{12}=4096$ \\
Conversion factor from CO inten- & $1.9\times 10^{20}$ \\
sity to H$_2$ column density $X_{CO}$ & mol cm$^{-2}$(K km/s)$^{-1}$ \\ 	
He/H ratio in the galactic gas & 0.1 \\		
		\end{tabular}
	\end{ruledtabular}
\end{table}

\subsection{\label{sec:res}Obtained emission characteristics}
\begin{figure}[H]
	\includegraphics[width=1\linewidth]{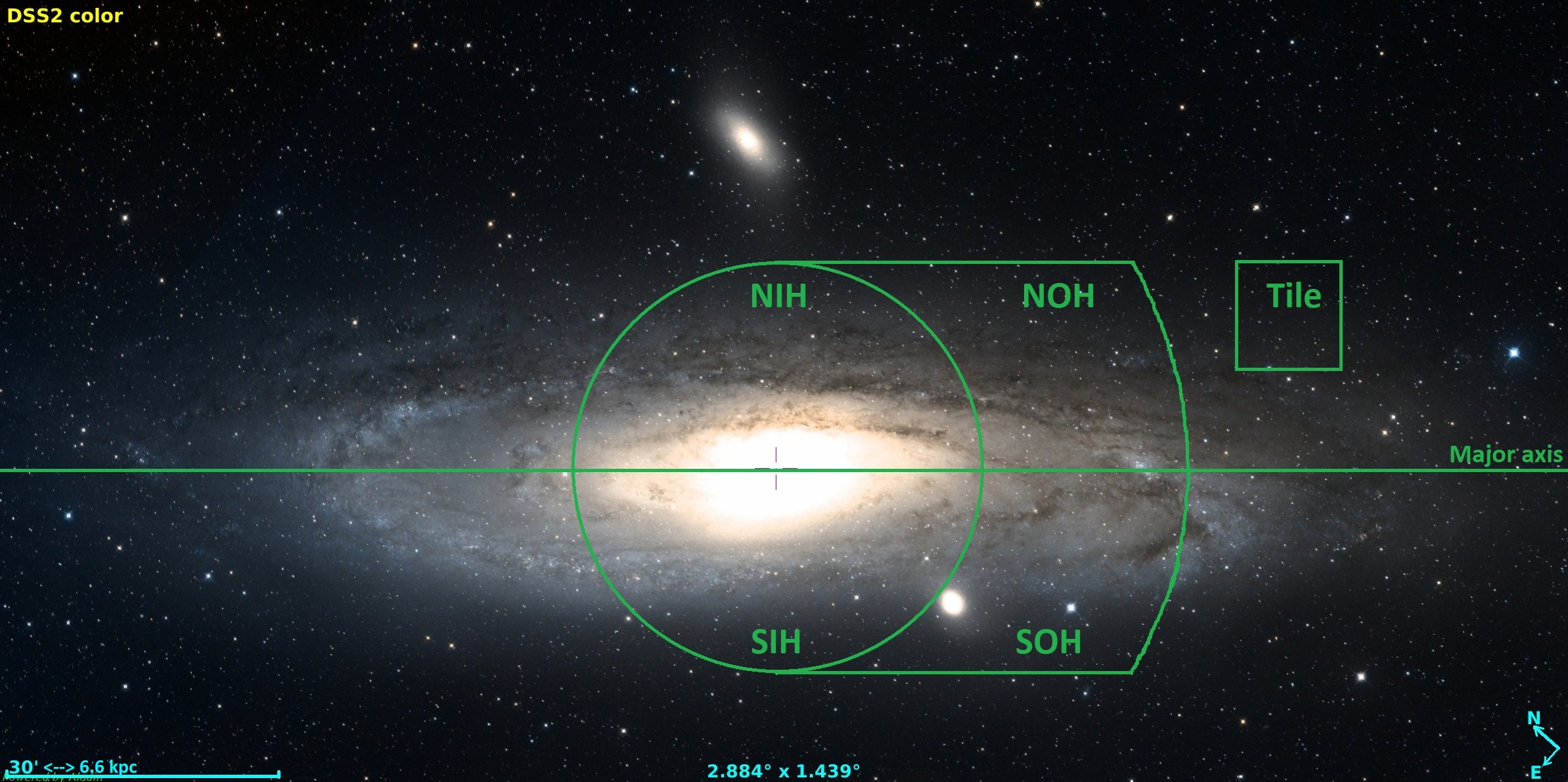}
	\caption{\label{fig:roi} The visual image of M31 (obtained from Aladin sky atlas) with the border lines of ROIs chosen for our analysis. The abbreviations have the following meanings: NIH -- northern inner halo ROI (i.e. the northern half of the inner halo ROI), SIH -- southern inner halo ROI, NOH -- northern outer halo ROI and SOH -- southern outer halo ROI. The square marked as "Tile" represents the uppermost and rightmost tile in Fig. \ref{fig:tiles}. The approximate angular size of the diffusion cylinder sky projection is $3.0\degree \times 1.2\degree$ (major $\times$ minor axes, base (MED) case). More details are in Sec. \ref{sec:as}. }
\end{figure}

Putting together all the model ingredients described in the previous subsection, the ICS emission maps were computed at discrete photon energies over the wide range 1 keV -- 10 GeV with the photon energy increment factor of 10. Let us start the analysis of obtained maps from the selection of relevant ROIs. Historically, as was described in Sec. \ref{sec:i}, two radial zones around M31 center were considered separately: inside and outside $R \approx 5~\text{kpc} \leftrightarrow 0.4\degree$. They seem to have different mechanisms of the gamma-ray emission generation. Based on these considerations, as a natural and intuitive choice, the following ROIs were constructed for the analysis here -- they are depicted in Fig. \ref{fig:roi}. The first ROI resembles the IH and is the disk with 5 kpc radius around the center. The second ROI is the annulus fragment restricted by $R$ = 5 kpc and $R$ = 10 kpc circles; and the lines, which are parallel to the major axis and are 5 kpc away from it. The major axis divides each of these ROIs into symmetric halves. One half is closer to the northwest, the other -- to the southeast. Let us briefly call them the northern and southern parts. These parts are designed for their intensity comparison in the study of ICS emission asymmetry. The second ROI is intended to represent the OH. Both ROIs are constructed in a way, that they do not approach too close to the boundary of the sky projection of the diffusion cylinder, where the intensity values might be imprecise according to the explanation above. Thus, it would not make sense to study the OH by, for example, the full 5--10 kpc annulus, since it goes far beyond the diffusion cylinder projection at the upper and bottom areas. Another potentially useful feature of the chosen ROIs, if to consider separately their halves, is that they both have the characteristic size of about 5 kpc, which approximately matches the resolution of Fermi-LAT in the angular measure at the energies 1--10 GeV \cite{Fermi}. One detail about the OH ROI is that, generally speaking, there are two such ROIs -- one on each side from the minor axis. However, it does not matter which of these two ROIs is considered, since the emission intensity map is symmetric about the minor axis. As was explained above, the asymmetry is anticipated only about the major (horizontal in Fig. \ref{fig:roi}) axis. And let us declare the exact definition of the asymmetry, which is used through this paper: this is the relative difference between the ICS emission intensities in the northern and southern halves of the designated ROIs, i.e. $I_N/I_S-1$. An intensity inside any region means the value averaged over all map pixels in that region.  

As the next step, let us discuss an important test of accuracy of the obtained maps. As was mentioned above, our constructed model of M31 implies, that the ICS emission asymmetry is caused by the ISRF anisotropy and inclination \textit{only}. Hence, the ICS emission maps must be perfectly symmetric about the major axis in the case of isotropic ISRF. And CMB ISRF component is very helpful in testing this implication, since the CMB emission is almost absolutely isotropic everywhere by its nature. CMB ICS emission component may have only a minor asymmetry of the second order due to the following peculiar mechanism. The primary asymmetry of the IR and optical ICS components implies the difference in ICS cooling rates for $e^\pm$, which reside in different hemispheres. In other words, $e^\pm$ from the brighter hemisphere radiate out their energy faster. And a faster cooling rate implies, in turn, a lower equilibrium concentration of emitting $e^\pm$. Therefore, we may expect a minor second order anisotropy with the opposite sign due to such a difference in cooling rates. And it may be notable mainly for the CMB ICS component, which lacks the first order anisotropy.

The asymmetry of CMB ICS emission component in all the computed parameter configurations at all photon energies in both ROIs does not exceed $\approx$ 1\% by absolute value. I attribute this small residual asymmetry to a combination of two effects: the described second order asymmetry and the finite map pixelization. A nearly perfect symmetry of the obtained CMB ICS emission maps confirms the overall correctness of the whole calculation algorithm. And the cited $\approx$ 1\% residual asymmetry represents and is accepted as a fair estimate of the uncertainty of all the asymmetry values calculated in the frame of our model in the selected ROIs. Indeed, this uncertainty does not include all systematical model uncertainties, like those related to a difference between the real M31 ISRF and the benchmark MW ISRF utilized here.

\begin{figure}[H]
	\includegraphics[width=0.497\linewidth]{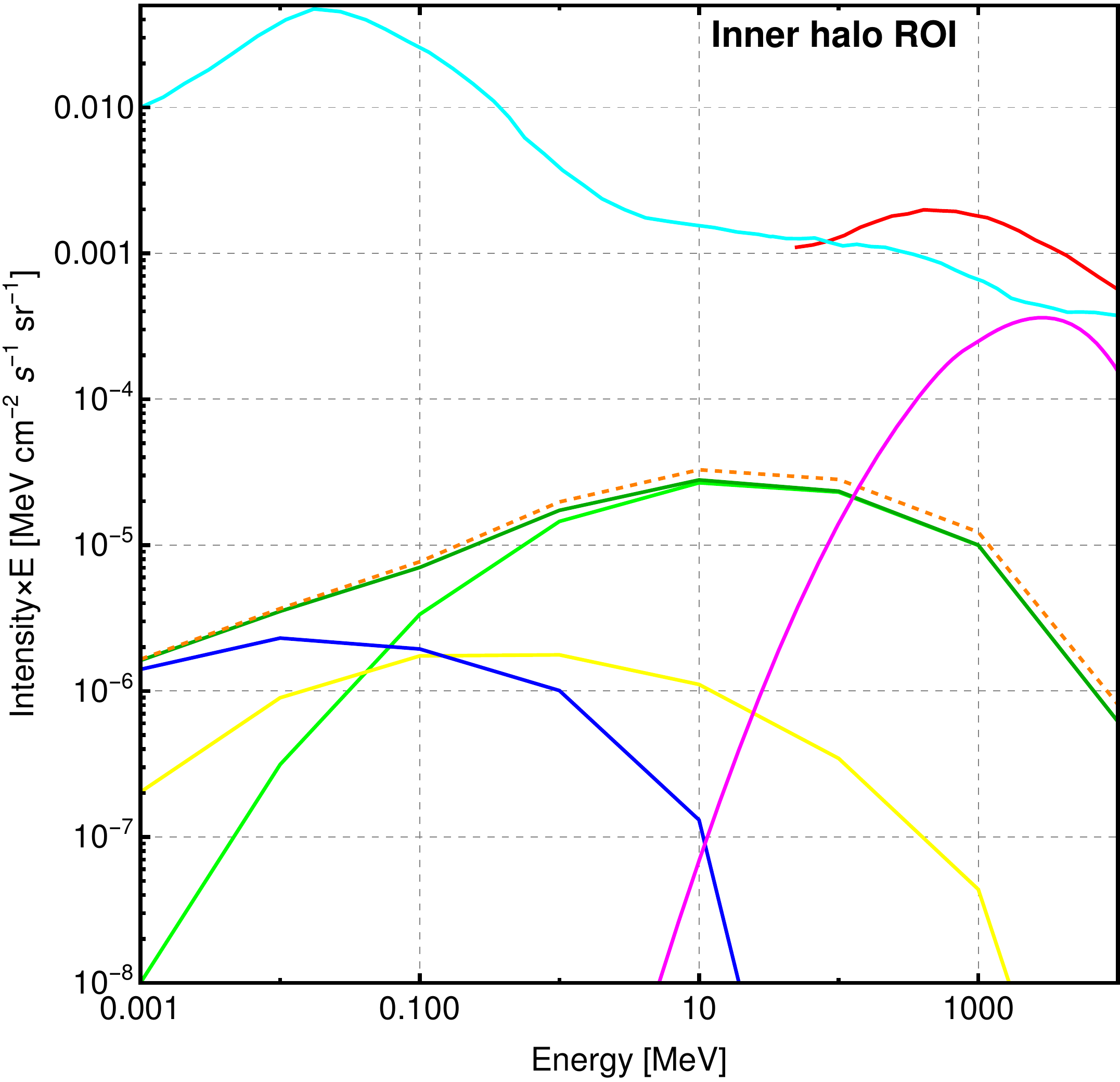}
	\includegraphics[width=0.497\linewidth]{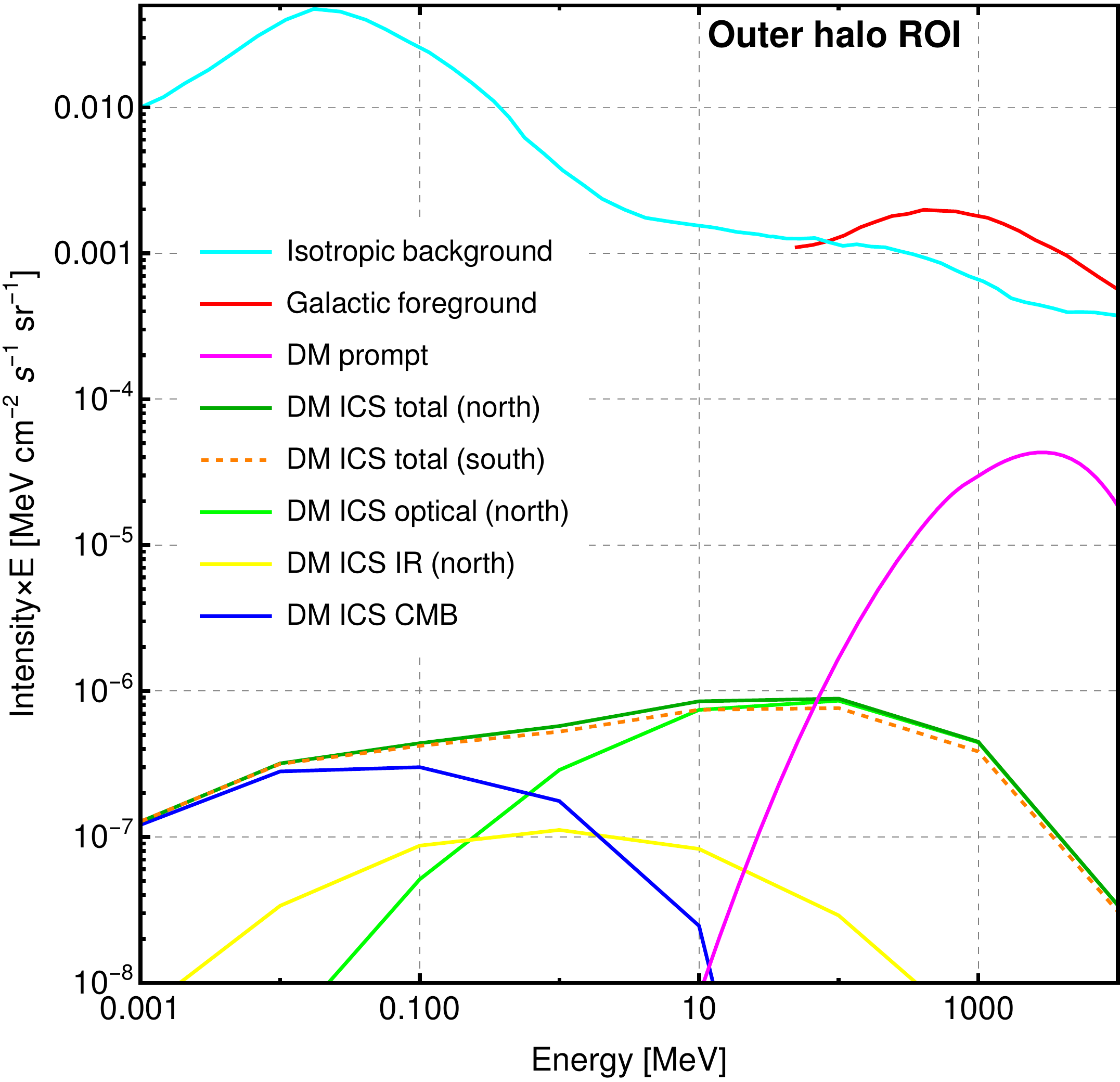}
	\caption{\label{fig:sp}The obtained ICS emission spectra in the IH ROI (\textit{left panel}) and in the OH ROI (\textit{right panel}) for our model $e^\pm$ population produced by the thermal WIMPs with $m_x$ = 60 GeV ($\chi\chi \rightarrow b\overline{b}$), which fit the OH gamma-ray emission according to \cite{2021PhRvD.103b3027K}. ICS emission from various ISRF components are shown individually. "North" and "south" mean respective ROI halves. The estimated spectrum of the isotropic background was taken from [\cite{2017ExA...tmp...24D}, Fig. 3]. The measured spectrum of the Galactic foreground at M31 location was taken from Fermi-LAT diffuse Galactic emission map \cite{Fermi-map}. More details are in Sec. \ref{sec:as}.}
\end{figure}

Now we can look at the obtained ICS emission spectra in the prepared ROIs. The spectra are shown in Fig. \ref{fig:sp} separately for IH and OH ROIs. For definiteness, the northern ROI halves were mainly used for Fig. \ref{fig:sp} (shown by the solid lines). Besides the total ICS emission spectra, the spectra from all three main ISRF components are also shown individually. According to the basic Eq. (\ref{eq:Eg}) above, the energies of photon before and after the scattering act are linearly proportional to each other. This fact is clearly reflected in Fig. \ref{fig:sp}: the ICS spectrum from CMB photons peaks at 0.01--0.1 MeV in the chosen units, from IR photons -- at 0.1--1 MeV and from optical photons (as well as the total) -- at 10--100 MeV. Overall, the contribution from optical photons dominates at energies above 0.1--1 MeV. The IH ROI is brighter than the OH ROI by more than an order of magnitude. This reflects a very steep radial dependence of DM annihilation rate. The spectral shape does not differ much between the IH and OH ROIs. The magenta lines represent the prompt (primary) gammas due to annihilation of the same WIMPs. The prompt gamma-ray emission maps were computed by the absolutely same GALPROP framework with only one difference: the medium emissivity integration along the line of sight was done up to $R \approx 200$ kpc, which approximately corresponds to the virial radius of M31 DM halo. The prompt emission component is expectedly brighter than the ICS component, although the latter peaks at significantly lower energies. The dashed lines mark the total ICS spectrum in the southern ROI halves. And in spite of the very large range over the vertical axis, we can clearly see the difference between the northern and southern ROI spectra, meaning that the predicted ICS emission asymmetry effect truly exists! And the asymmetry has different signs in the IH and OH ROIs. In order to have some connection with the real sky, Fig. \ref{fig:sp} also contains the approximate measured spectra of the isotropic background and Galactic foreground at M31 location. The former does not continue below 50 MeV due to a scarcity of data. Another relevant aspect is the actual observed intensity of M31. Concerning the IH ROI, according to \cite{2023ApJ...945L..22X} the point source (if to spread it over the whole IH) in M31 center produces the intensity, which is approximately the same as the Galactic foreground shown by the red line in Fig. \ref{fig:sp}. For the OH, a good estimate of M31 intensity would be the prompt DM emission spectrum shown by the magenta line, since DM model used here fits all the OH emission according to \cite{2021PhRvD.103b3027K}.
\onecolumngrid

\begin{figure}[H]
	\includegraphics[width=1\linewidth]{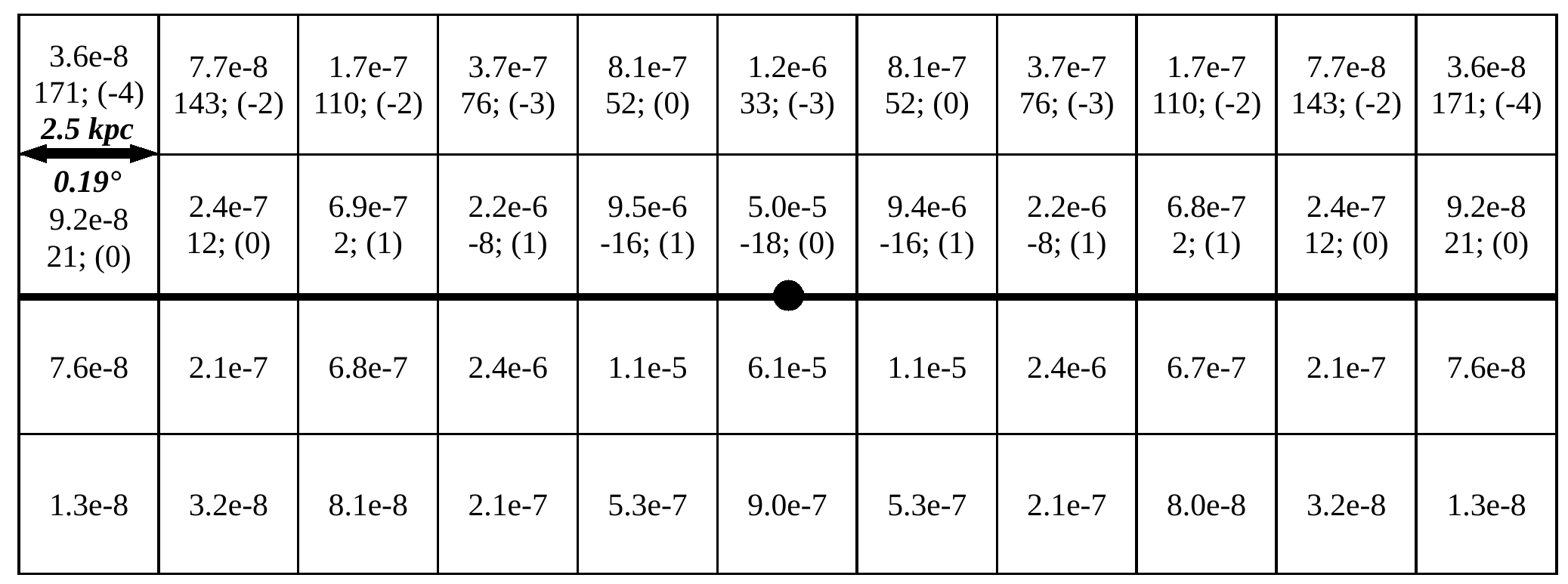}
	\caption{\label{fig:tiles}The ICS emission intensity and asymmetry map at $E_\gamma$ = 1 GeV for the base parameter configuration by means of the square tiling. The thick line marks M31 major axis, the thick dot -- the galactic center. Each tile contains the following values: the average intensity (multiplied by the photon energy) in the units [MeV cm$^{-2}$ s$^{-1}$ sr$^{-1}$] (with "e" meaning $\times 10\string^$), the total asymmetry defined here as the relative intensity difference in \% between the upper row tiles and their symmetric lower counterparts, and the asymmetry of CMB ICS component (in brackets). The upward direction corresponds to northwest. }
\end{figure}
\begin{table}[H]
	\caption{\label{tab:as}The ICS emission asymmetry [\%] dependence on the photon energy and model parameters configuration. The asymmetry values are provided for both IH ROI and OH ROI (separated by ";", "[,]" means a range of values). }
	\begin{ruledtabular}
		\centering
		\begin{tabular}{ccccccc}
			Parameter configuration & 0.1 MeV & 1 MeV & 10 MeV & 100 MeV & 1 GeV & 10 (100) GeV \\
			\Xhline{1pt}
			Base: $m_x$ = 60 GeV, $\chi\chi \rightarrow b\overline{b}$, &  &  &  &  &  &  \\
			MAX DM density, MED MF/prop.                                & --7; 4 & --11; 8 & --13; 13 & --14; 15 & --16; 16 & --21; 10 \\
			\hline
			Base with DM prompt gammas added. & --7; 4 & --11; 8 & --13; 13 & --10; 5 & --1; 0 & --1; 0 \\
			\hline
			Base with misc parameter variations:              & [--7, --2]; & [--11, --3]; & [--13, --5]; & [--14, --8]; & [--16, --10]; & [--21, --17]; \\        
			$i = [70\degree, 77\degree]$, [MIN, MAX] MF/prop. & [3, 8] & [6, 17] & [10, 24] & [11, 26] & [12, 27] & [6, 22] \\
			\hline
			Base with misc parameter variations  & [--7, --2]; & [--11, --3]; & [--13, --5]; & [--10, --5]; & --1; & --1; \\        
			and DM prompt gammas added.          & [3, 8] & [6, 17] & [10, 24] & [4, 8] & 0 & 0 \\
			\hline
			Base except $m_x$ = 600 GeV. & --4; 1 & --7; 3 & --10; 8 & --12; 14 & --14; 18 & --15; 16 (--17; 13) \\
			\hline                           
			Base except $\chi\chi \rightarrow \tau^+\tau^-$. & --3; 0 & --5; 2 & --9; 10 & --12; 18 & --14; 18 & --18; 15 \\
			\hline                                                 
			Base except MIN DM density. & --3; 3 & --5; 6 & --7; 8 & --9; 7 & --12; 6 & --19; 0 \\
		\end{tabular}
	\end{ruledtabular}
\end{table}
\twocolumngrid

Then let us analyze the morphology of the ICS emission.  In order to visualize the intensity and asymmetry map, I tiled the sky projection of the diffusion cylinder at the representative photon energy of 1 GeV by the squares with 2.5 kpc sides symmetrically about M31 center -- see Fig. \ref{fig:tiles}. The thick line there denotes the major axis, the thick dot -- M31 center. Also, Fig. \ref{fig:roi} illustrates the position of the uppermost rightmost tile on the sky. The specific size of 2.5 kpc was chosen due to the following reasons. It provides a convenient tiling of the diffusion cylinder projection without too close approach to the boundary and with the sufficient number of map pixels ($\approx$ 170) inside every tile for mitigating pixel "noise". Also, 2.5 kpc corresponds to the angular resolution, which may be achievable in the near future with new gamma-ray telescopes (see, e.g., [\cite{2017ExA...tmp...24D}, Fig. 19]). The tiles contain their ICS emission intensity value, and the asymmetry values of the total intensity and its component from CMB ISRF (in brackets). The asymmetry in Fig. \ref{fig:tiles} means the relative intensity difference between the tiles in the rows above the major axis and their symmetric counterparts below.

Looking at Fig. \ref{fig:tiles} we may notice, first of all, that the intensity distribution is almost perfectly symmetric about the minor galactic axis. This matches expectations from our model of M31. The intensity steeply decreases with the central distance. Considering the asymmetry distribution, we see a peculiar picture. Naively, we might expect the asymmetry to have the same sign over the whole hemisphere, i.e., when one hemisphere is uniformly brighter than the other one. However, the asymmetry is very non-uniform: it is negative close to the center in the bulge region and positive elsewhere. The asymmetry goes down to --18\% near the center and rises up to 170\% in the corners. The asymmetry of CMB component serves as an indicator of the correctness of the calculation algorithm. As was explained above, the CMB component must be nearly symmetric. It is indeed so in the second row from the top. However, the first row demonstrates some minor systematically negative CMB asymmetry at the level of few \%, which slightly exceeds our uncertainty estimate of 1\%. This asymmetry can be attributed to two possible natural reasons. One is the second order asymmetry due to the difference in cooling rates described above. Indeed, the primary asymmetry in the first row tiles is very large, therefore $e^\pm$ cool down much faster there. And the second reason may be geometrical: the upper (northwest) part of M31 disk is further from us than the lower (southeast) part according to \cite{2010A&A...509A..91T}. Then the difference in distances to the emitting regions corresponding to the tiles in the first and fourth rows is larger than that to the tiles in the second and third rows (see also Fig. \ref{fig:ics}). Therefore, the tiles in the first row may look fainter than their counterparts in the fourth row, but for the second row, this effect is still not pronounced. Thus, we may conclude again, that our model reasonably reproduces the symmetry of CMB component.

Now let us try to deduce an intuitive physical interpretation of the obtained non-uniform asymmetry distribution. Fig. \ref{fig:ics} serves for this purpose. In general, we may expect, that the biggest ICS emissivity is reached in the places with the highest ISRF density, i.e. inside M31 bulge and disk. Let us choose four trial points (1--4 at Fig. \ref{fig:ics}) there in order to study the geometry of ICS process. Let us assume that the near vicinity of these points produces a defining contribution to the emission intensity integrated along the whole line of sight inside the diffusion cylinder. The points 1 and 4 represent pairs of locations outside the bulge, the points 2 and 3 -- inside. $\vec{V_e}$ is the velocity of emitting $e^\pm$ before scattering. And the photon symbols represent the supposed prevailing momentum direction of ISRF target photons before scattering. Thus, $\theta_{i}$ are the angles between the initial momenta of photons and $e^\pm$. According to Eq. (\ref{eq:Eg}) above, the energy of upscattered photon grows monotonically with $\theta_{i}$. It means that the energy transfer between $e^\pm$ and a photon happens most efficiently in head-on collisions and least efficiently, when $e^\pm$ chases a photon moving in nearly the same direction. Therefore, the ICS emission intensity increases with $\theta_{i}$, if to consider the latter as the effective average angle, which is defined by the prevailing direction of photon motion. I attempted to guess such directions at Fig. \ref{fig:ics}. Considering points 1 and 4, the ISRF intensity in the direction along the galactic plane should be much higher than across. And the intensity in the direction outward from the nucleus is probably slightly higher than toward, at least because ISRF density is generally expected to decrease with the radius. These intuitive considerations led to the marked prevailing photon momenta directions at points 1 and 4. $\theta_{i1} > \theta_{i4}$, hence the upper (northern) hemisphere outside the bulge region must be brighter than the lower (southern) one, which reconciles with Fig. \ref{fig:tiles}. But when our line of sight dives into the bulge, the situation is less obvious -- see points 2 and 3. The bulge has a spheroidal shape, and ISRF is expected to be rather isotropic inside. Hence, the bulge by itself does not seem to provide any significantly prevailing direction of photon momenta. The former may be produced by the photons from the disk parts, which are closest to points 2 and 3, as shown in Fig. \ref{fig:ics}. Then $\theta_{i2} < \theta_{i3}$, which explains the negative asymmetry in the bulge region according to Fig. \ref{fig:tiles}. This is a qualitative interpretation of the peculiarities of obtained asymmetry distribution best to the author's imagination. However, other mechanisms may indeed work too. Thus, we have also keep in mind, that the medium ICS emissivity also depends on $e^\pm$ concentration, which has a steep radial profile defined majorly by DM density profile. And the interpretation above considered, for simplicity, just the plane section of the system, although the volumic picture does not seem to introduce anything essentially else.

Finally, let us analyze the asymmetry dependence on the photon energy and other model parameters. Table \ref{tab:as} serves for this purpose. It lists the asymmetry values starting from $E_\gamma$ = 0.1 MeV for both IH and OH ROIs (the same ROIs as were used for Fig. \ref{fig:sp}). Below $E_\gamma \sim$ 0.1 MeV the asymmetry vanishes, because the CMB ISRF component dominates in the production of ICS emission and makes it symmetric at those energies, as can be seen in Fig. \ref{fig:sp}. We can see from the table, that the absolute value of asymmetry generally grows with energy for all the computed parameter configurations. At low energies this trend is caused again by a decrease of CMB ICS contribution with the energy increase. Considering the IH ROI, the growth of asymmetry is monotonous there over the whole energy range. In the OH ROI the asymmetry reaches its maximum typically around $E_\gamma \sim$ 1 GeV and then decreases. The first row of the table represents the base parameter configuration, which was described in Sec. \ref{sec:mod} and fits the OH emission according to \cite{2021PhRvD.103b3027K}. We may note for this case, that the asymmetry is quite substantial: it reaches --21\% in the IH ROI and 16\% in the OH ROI. The second row reflects the same parameter configuration, but when the prompt gamma-ray emission due to WIMP annihilation is added on top of the ICS emission. The prompt emission spectrum is relatively narrow: as can be seen in Fig. \ref{fig:sp}, it is narrower than the ICS emission spectrum. The prompt emission reaches a significant contribution and begins to influence the asymmetry around $E_\gamma \sim$ 100 MeV, and then vanishes the asymmetry completely at $E_\gamma \gtrsim 1$ GeV due to a full dominance over the ICS emission at those energies.

The third and fourth rows of Table \ref{tab:as} illustrate the ranges of asymmetry values (around the base scenario) due to the uncertainties in M31 inclination and MF/prop. model parameters. The inclination angle ($i$) value has key importance for the ICS emission asymmetry effect. For the base scenario $i = 74\degree$ was assumed as the average value among various determinations cited in \cite{2021ApJ...920...84L}. Those determinations obtained values ranging from 70$\degree$ to 77$\degree$. And I computed the asymmetry values for these boundary inclination values too in order to study the respective uncertainties. MF/prop. parameters were also varied together with the inclination, although the former are expected to play a secondary role. The ranges of possible values of MF/prop. parameters were taken from [\cite{2022PhRvD.106b3023E}, Table III]. As can be seen in Table \ref{tab:as}, the resulting uncertainty ranges of the asymmetry are quite substantial, especially in the OH ROI. The IH ROI did not show a clear correlation between the asymmetry and the varied model parameter values. A low sensitivity to the inclination value in the IH ROI could be due to the fact, that the bulge and ISRF there have quasi-spherical geometry.

The next rows of the table show the asymmetry for alternative DM models: heavier (by an order of magnitude) WIMPs with $m_x$ = 600 GeV, another important annihilation channel $\chi\chi \rightarrow \tau^+\tau^-$ and MIN DM density profile (from [\cite{2022PhRvD.106b3023E}, Sec. IIIB]). In each of these three cases, only the mentioned model element was changed, other parameters correspond to the base configuration. The listed DM models do not have direct relevance for the fit of outer gamma-ray halo. They were computed for a somewhat different purpose -- to study the asymmetry dependence on the properties of $e^\pm$ population. In a broader view, WIMP mass (i.e., the rest energy) can be considered as the upper limit for the energy of $e^\pm$. WIMP annihilation channel defines the shape of $e^\pm$ energy spectrum (at the injection). DM density profile defines the degree of concentration of $e^\pm$ source. Hence, by varying these DM model parameters, we may explore to some extent an overall possible variation of the asymmetry due to a variation of the properties of $e^\pm$ population. The thermal WIMPs much lighter than 60 GeV were not considered, because they are quite robustly excluded by various indirect searches and, hence, are not interesting. The fifth row of Table \ref{tab:as} reflects the case of heavier WIMPs. In this case we may note slightly smaller asymmetry at low photon energies and a confident propagation of the asymmetry to very high energies $E_\gamma \sim 100$ GeV. This can be explained by an overall shift of $e^\pm$ population toward higher energies. The noted persistence of the asymmetry at $E_\gamma \sim 100$ GeV can be considered as an indication of possibility of the asymmetry existence at an arbitrarily high energy, as far as $e^\pm$ exist with that energy. The sixth row shows the case of $\chi\chi \rightarrow \tau^+\tau^-$: at low photon energies it shows appreciably lower asymmetry with respect to the base configuration. And the last row of the table reflects the case of very different DM density profile: MIN, i.e. the cored profile, which is opposite to the MAX cusped profile. As was discussed in [\cite{2022PhRvD.106b3023E}, Sec. IIIB], MIN and MAX profiles approximately enclose the range of possible DM densities in the central region (inner several kpc) of M31 DM halo. MIN profile demonstrates significantly lower asymmetry values at all the energies in both ROIs in comparison with the base case. Thus, the degree of concentration, i.e. the radial steepness of $e^\pm$ source distribution, substantially influences the asymmetry. The zero value of the asymmetry in the OH ROI at $E_\gamma = 10$ GeV does not necessarily mean full symmetry: likely, it is just the result of superposition of negative and positive asymmetries in different parts of the OH ROI.

Looking at Table \ref{tab:as} overall, we can find out, that the negative asymmetry in the IH ROI reaches at most --21\%, while the positive asymmetry in the OH ROI -- +27\%. Indeed, the computed trial cases do not provide full and smooth coverage of the whole parameter space. But they outline well the characteristic asymmetry values, which we may anticipate from some arbitrary $e^\pm$ population. Thus, concluding broadly, the ICS emission from any reasonable $e^\pm$ population is expected to possess an asymmetry at the level of 10\%--20\% (at least at some energies) in the selected ROIs.

\subsection{\label{sec:obs}Observational prospects}
Although this paper aims mainly at the theoretical modeling of the ICS emission asymmetry effect, this subsection discusses some general considerations of the effect's observability. Let us go back to Fig. \ref{fig:sp}, which reflects the base thermal WIMP scenario for the outer gamma-ray halo fit. Considering the ICS emission intensity averaged over IH and OH ROIs, we see that, unfortunately, it is smaller than the background and foreground emission intensities by more than one order of magnitude at all photon energies in both ROIs. Hence, it would be very challenging to separate even just the ICS component itself from other emission components in the direction of M31. The asymmetry, i.e. the intensity difference between the ROI halves, is even smaller: it can be seen as the difference between the dark-green and dashed lines. In comparison with the backgrounds/foregrounds, this difference is tiny at all. However, one may consider other regions, where the asymmetry values are larger: according to Fig. \ref{fig:tiles}, the asymmetry exceeds 100\% in the locations far from M31 center, i.e., the intensity ratio exceeds factor of 2 there. But the problem for such locations is a very small intensity amplitude, since the latter steeply decreases with the distance from center. Hence, changing the ROIs does not seem to provide an obvious sensitivity gain. Thus, at first glance, the overall picture appears to be quite pessimistic in the sense; that the predicted asymmetry effect indeed exists, but mainly theoretically, since its observational detection requires a virtually unlimited instrumental sensitivity. 

At the same time, there are some reasons for optimism. One reason is that, according to Fig. \ref{fig:sp}, the prompt and ICS emission spectra peak at significantly different energies. This circumstance and a relative narrowness of the prompt spectrum loosen an observational degeneracy between these two components: the bright prompt signal does not "interfere" much with the weaker ICS signal and its asymmetry at $E_\gamma \lesssim 100$ MeV. The latter energy range can be optimal for a search of the asymmetry also due to another reason: photon fluxes are relatively high there, since they steeply decrease with energy. At higher gamma-ray energies $E_\gamma \gtrsim 1$ GeV precise intensity measurements are very limited due to small photon counts and, hence, large statistical errors. But at lower energies photon counts are large (for the same detector area); which may allow, in principle, precise intensity measurements and, hence, separation of faint emission components. Another challenge is the angular resolution: that achievable now by Fermi-LAT at low energies is insufficient. However, in general, we may anticipate significant progress in both the sensitivity and angular resolution of gamma-ray telescopes in the future. The relevant future missions, which are expected to have a good sensitivity at low energies, include e-ASTROGAM \cite{2017ExA...tmp...24D}, AMEGO \cite{2019BAAS...51g.245M}, HERD \cite{2022icrc.confE.651F} and GAMMA-400 \cite{2018PAN....81..373E,2020JCAP...11..049E}.

So far, the discussion has been tied solely to the specific thermal WIMP, which is needed for the fit of OH. However, according to [\cite{2021PhRvD.103b3027K}, Table II], the fit may require higher than thermal cross sections even for the same (high) J-factor values (the exact definition of J-factor notion is given by Eq. (7) there). The required cross section value is highly uncertain due to the complicated systematic uncertainties described there. The ICS emission intensity is linearly proportional to the cross section. Hence, in the case of higher cross sections, the ICS signal would be proportionally brighter and, therefore, easier to detect over backgrounds.

As was emphasized in Sec. \ref{sec:i}, the observed gamma-ray emission in M31, speaking generally, must not necessarily originate from DM. The emission may come from a population of CRs through both the leptonic and hadronic mechanisms. If the leptonic component, i.e. the ICS emission from CR $e^\pm$, is non-zero, then the asymmetry of the total emission would indeed be non-zero too. The asymmetry of ICS emission from CR $e^\pm$ is even easier to detect in comparison with DM $e^\pm$ case; because in the latter case the ICS emission has a secondary role, and the asymmetry is present only inside a relatively narrow energy range. In the CR case there is no prompt gamma-ray component, i.e. the ICS component is primary, and only a potential hadronic component may "interfere" with the ICS emission of interest. Also in this case the ICS emission spectrum and its asymmetry do not have a steep cut-off near WIMP rest energy -- at $E_\gamma \sim 10$ GeV in our case. The spectrum from CR $e^\pm$ would be much wider over the energy. The asymmetry is naturally expected to be the most prominent in the case, when all the observed emission comes from CR $e^\pm$. Summarizing, detection of the asymmetry effect is a very challenging task, but potentially not hopeless for future highly sensitive gamma-ray telescopes. 

\section{\label{sec:sec}Estimation of DM $\mathbf{e^\pm}$ ICS and bremsstrahlung emission contributions}

This section is dedicated to a somewhat different aspect of DM interpretation of the outer gamma-ray halo. Best to the author's understanding, the work \cite{2021PhRvD.103b3027K} took into account only the prompt emission component due to DM annihilation for the fit of OH emission and neglected by the secondary components -- ICS and bremsstrahlung from DM $e^\pm$. This section aims to check the validity of such approximation by estimating the emission fluxes from these secondary contributions and comparing them with the prompt emission flux. This estimation was done for the same thermal WIMP with $m_x$ = 60 GeV annihilating to $b\bar{b}$, which fits the OH. The secondary emission fluxes were calculated for the annular region on the sky, which was used in \cite{2021PhRvD.103b3027K} for the fitting procedure and called "spherical halo (SH)" there. This annulus is centered at M31 center, and has angular radii 0.4$\degree$ and 8.5$\degree$. For definiteness, I consider here the case "I" according to the classification in \cite{2021PhRvD.103b3027K}; i.e., when annihilating WIMPs in both MW and M31 form the signal along M31 line of sight, and there is no effect of absorption of MW DM signal by isotropic emission template. The secondary emission fluxes were computed here individually for MW and M31. Regarding DM density profile, it is the same here as in the previous section (Einasto MAX). The profile parameters for MW were obtained according to the recipe in \cite{2021PhRvD.103b3027K}. And MED MF/prop. configuration was employed.

Let us start the discussion from the ICS component of secondary emissions. M31 ICS emission intensity maps were already calculated by GALPROP in the frame of modeling conducted in the previous section. Here we need in the emission flux from the annular region cited above for a convenient comparison with the prompt fluxes in \cite{2021PhRvD.103b3027K}. This region encloses the whole M31 diffusion zone sky projection except the central disk with 0.4$\degree$ radius. MW ICS emission intensity maps were calculated by the same way with DM density and MF/prop. parameters adapted for MW. MW ICS emission spreads over the whole sky, indeed, since we look at the sky from the inside of MW diffusion cylinder (meanwhile, \cite{2023PDU....3901157D} has a relevance in this context). The extracted ICS emission fluxes from the annular ROI at three representative energies are written out in Table \ref{tab:sec}.

Another potentially relevant emission mechanism is bremsstrahlung from DM $e^\pm$. Bremsstrahlung emission maps were computed by GALPROP too. This calculation requires knowledge of the interstellar gas distribution in both galaxies. GALPROP conveniently provides the relevant gas distributions for MW. This allows to compute quite detailed all-sky bremsstrahlung intensity maps. For M31 such detailed gas maps do not exist, and MW gas maps were substituted. Such a rough approximation is acceptable due to the following reasons. First, contrary to MW, we do not need in the detailed intensity map of M31. Instead, we need in just the total flux from almost the whole galaxy (without its central part). The total bremsstrahlung flux from the whole galaxy is presumably proportional to the total galactic gas mass. MW and M31 have very similar total gas mass values -- they are provided in Table \ref{tab:lgm}. Hence, MW gas distributions are expected to provide a reasonable estimate of the total bremsstrahlung flux from M31. Also, Table \ref{tab:gp} displays the values of certain parameters used in the bremsstrahlung calculation procedure. The resulting bremsstrahlung fluxes are displayed in Table \ref{tab:sec}.

Now we can compare the obtained secondary emission fluxes with the prompt ones. The latter were extracted from [\cite{2021PhRvD.103b3027K}, Fig. 2 (right), the whole SH] and are written out above the horizontal line in Table \ref{tab:sec}. In general, we can note that the prompt emission flux is much larger than the ICS one, and the latter in turn is much larger than the bremsstrahlung flux. Thus, the latter plays the least role among three components and, by the way, would not create much of a nuisance background for detection of the ICS emission asymmetry. All MW emissions are much brighter than M31 ones. \cite{2021PhRvD.103b3027K} used the photon energy range starting from 1 GeV for the fit. We see from Table \ref{tab:sec}, that all the secondary components are negligible in comparison with both MW and M31 prompt components at $E_\gamma \gtrsim 1$ GeV. Thus, it is valid to neglect by secondary emissions in the fit at those energies. However, at $E_\gamma = 0.1$ GeV we can see, that MW secondary flux becomes comparable to M31 prompt flux. Hence, if one would employ lower energies $E_\gamma \lesssim 1$ GeV for the fit, then it is necessary to take into account at least MW ICS component (if the scheme "I" is realized, when both galaxies are included).

And there is another important caveat, which concerns the annihilation cross section value. So far we have discussed the case of thermal cross section, which is the most natural and approximately corresponds to the parameter configuration described by the first row of Table II in \cite{2021PhRvD.103b3027K}. However, other rows of this table represent alternative viable parameter configurations: they have less extreme substructure boosts, but require more exotic DM with high annihilation cross sections. Lowering the boost factor would not decrease the secondary emission fluxes significantly, since the boost factor works mainly on the largest radii $R \sim 100$ kpc and does not affect the annihilation rate all that much inside the diffusion cylinder, i.e. at $R \lesssim 20$ kpc. But increasing the cross section would linearly increase the secondary fluxes. Hence, for lower boost factor cases, the secondary emission fluxes would be higher and contribute more to the total flux. Thus, one may need to take into account certain secondary components, if the cross sections higher than thermal are employed for the fit.

Speaking very generally, there is another -- the fourth -- gamma-ray emission type due to annihilating WIMPs: the emission from decaying pions, which are produced by DM proton and antiproton collisions with an interstellar gas. However, hadronic emissions are beyond the scope of this work.

\begin{table}[h]
	\caption{\label{tab:sec}The emission fluxes in the units [MeV cm$^{-2}$ s$^{-1}$] of various components due to DM annihilation in MW and M31 from the annular ROI centered at M31 center with angular radii 0.4$\degree$ and 8.5$\degree$. The prompt emission fluxes above the horizontal line are taken from \cite{2021PhRvD.103b3027K}. DM parameters correspond to the OH fit. }
	\begin{ruledtabular}
		\centering
		\begin{tabular}{cccc}
			& Flux at & Flux at & Flux at \\
			Emission component & 0.1 GeV & 1 GeV & 10 GeV \\
			\Xhline{1pt} \noalign{\vskip 1pt}
			Total observed =     & $3\times 10^{-7}$ & $6\times 10^{-6}$ & $4\times 10^{-6}$ \\
			MW prompt +          & $3\times 10^{-7}$ & $5\times 10^{-6}$ & $3\times 10^{-6}$ \\
			M31 prompt           & $4\times 10^{-8}$ & $7\times 10^{-7}$ & $4\times 10^{-7}$ \\	
			\hline \noalign{\vskip 1pt}
			MW ICS +             & $2\times 10^{-8}$ & $1\times 10^{-8}$ & $7\times 10^{-10}$ \\
			MW bremsstrahlung =  & $6\times 10^{-9}$ & $4\times 10^{-9}$ & $2\times 10^{-10}$ \\
			MW secondary         & $3\times 10^{-8}$ & $1\times 10^{-8}$ & $8\times 10^{-10}$ \\
			M31 ICS +            & $2\times 10^{-10}$ & $8\times 10^{-11}$ & $6\times 10^{-12}$ \\
			M31 bremsstrahlung = & $7\times 10^{-12}$ & $6\times 10^{-12}$ & $3\times 10^{-13}$ \\
			M31 secondary        & $2\times 10^{-10}$ & $9\times 10^{-11}$ & $7\times 10^{-12}$ \\
		\end{tabular}
	\end{ruledtabular}
\end{table}	

\section{\label{sec:dm}Interrelation between DM radio constraints and fit of the outer halo}

This section aims to discuss another aspect, which is very relevant for DM explanation of the outer gamma-ray halo emission: how WIMP parameter values, which are needed for the fit, relate to the constraints obtained from radio observational data on M31. DM $e^\pm$ inevitably produce synchrotron emission too in the galactic MF in the radio band. And radio observations provide a powerful tool for constraining the properties of any $e^\pm$ population. This aspect was already discussed briefly in [\cite{2022PhRvD.106b3023E}, Sec. VI]. Here more details are revealed.

\begin{figure}[h]
	\includegraphics[width=1\linewidth]{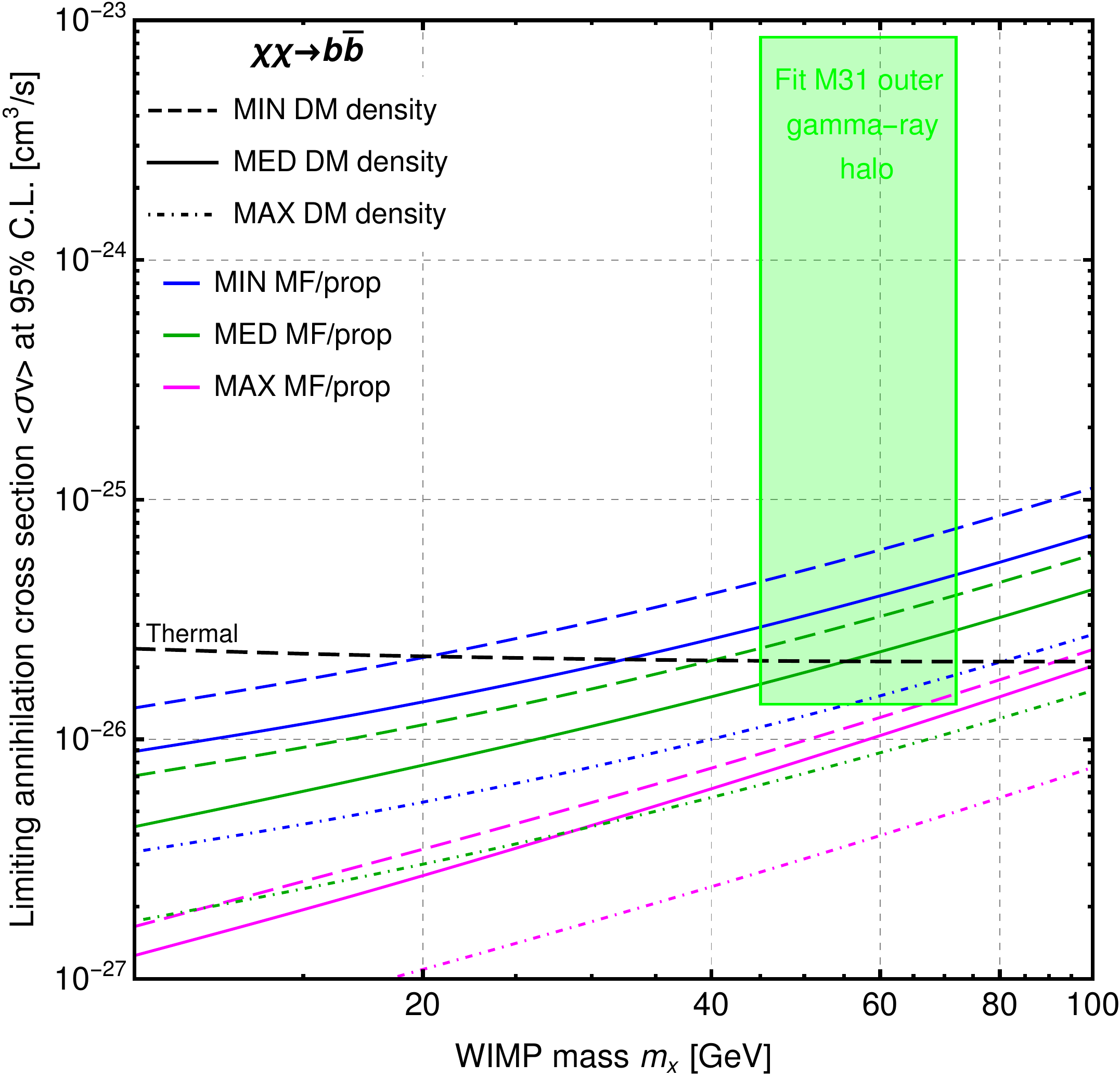}
	\caption{\label{fig:sv}WIMP parameter plane. The green rectangle shows the parameter region, which fits the outer gamma-ray halo according to [\cite{2021PhRvD.103b3027K}, Table II] (both I and II scenarios, Einasto profile). The lines represent the parameter constraints from radio observations at 95\% confidence level from \cite{2022PhRvD.106b3023E} for various commented DM density and MF/prop. parameter configurations. The black dashed line reflects the thermal cross section (taken from \cite{2020JCAP...08..011S,sv}). More details are in Sec. \ref{sec:dm}.}
\end{figure}

The main difficulty here is the fact, that the radio and gamma-ray data are being compared in essentially non-overlapping spatial regions of M31. Thus, \cite{2021PhRvD.103b3027K} conducted the fit avoiding the IH region, i.e., the fit was done over the radial distances $R \gtrsim 5$ kpc $\leftrightarrow$ 0.4$\degree$. However, the radio constraints \cite{2022PhRvD.106b3023E} were derived oppositely inside the IH region using $R \lesssim 3$ kpc (outside, these constraints are weak). DM density profile is determined quite precisely in the OH region -- this can be seen in, e.g., [\cite{2021PhRvD.103b3027K}, Fig. 3]. However, this can not be said about the IH region -- DM density is constrained rather poorly there (e.g., \cite{2012A&A...546A...4T}). [\cite{2022PhRvD.106b3023E}, Fig. 1] demonstrates the range of possible DM densities in the IH region. This wide range translates into a large uncertainty of WIMP parameter constraints derived from the IH. Fig. \ref{fig:sv} shows WIMP parameter plane (the annihilation cross section vs. mass plane) with the constraints derived in \cite{2022PhRvD.106b3023E} from the radio non-thermal emission maps of M31 bulge region and the parameter region required for the outer gamma-ray halo fit. The fitting parameter region (displayed by the green rectangle) was taken from \cite{2021PhRvD.103b3027K}, the cross section values are from the last two columns of Table II there. Both scenarios I and II were included here. These cross section values were taken without the uncertainties related to those in J-factor values (columns 7--9 there). The exclusion limits in Fig. \ref{fig:sv} came from [\cite{2022PhRvD.106b3023E}, Fig. 8(left)]. Their line style encodes the type of DM density profile, line color -- MF/prop. configuration according to \cite{2022PhRvD.106b3023E}. Strictly speaking, we must be comparing WIMP parameters in the radio and gamma-ray bands for \textit{exactly} similar DM density profiles in order to draw meaningful inferences, because every profile requires its own fitting cross section values. The exact similarity holds only for the MAX profile among three profile cases depicted. MED and MIN profiles may require slightly different cross sections for the outer gamma-ray halo fit, i.e. the top and bottom green rectangle borders at Fig. \ref{fig:sv} for them may be shifted with respect to the MAX profile case depicted. It is important to keep this caveat in mind interpreting this figure.

Another relevant circumstance is the difference in substructure boost factor model between DM density profiles used in \cite{2022PhRvD.106b3023E} and \cite{2021PhRvD.103b3027K}. However, this difference does not have a significant impact in our case. The reason is that the radio constraints were derived employing mainly only the central region of M31 with $R \lesssim 3$ kpc, where the role of substructures is very small due to their tidal disruption. Thus, it was calculated that the cross section radio limits differ just by 4\%--6\% between the cases with and without the substructure boost of MAX DM density profile (the boost factor radial profile was from \cite{2010PhRvD..81d3532K} for this case). Hence, the radio limits are quite universal with respect to the assumed boost factor model, and they can be compared directly (in \textit{this} aspect) to any cases outlined in [\cite{2021PhRvD.103b3027K}, Table II] within $\approx$ 10\% precision.

We see from Fig. \ref{fig:sv}, that in the case of MAX DM density profile, which provides the "pure" comparison, the radio limits exclude the gamma-ray fit nearly completely for all MF/prop. configurations. However, this can not be said about other profile cases. Low DM density and MF strength values in the central region of M31 allow some part of the green parameter region. But even the weakest MIN limit (the dashed blue line) allows, in fact, only a very small ($<$ 1\% by relative area in the linear measure) part of the gamma-ray fitting region. We can also notice, that the case II, when all the gamma-ray emission in OH comes from M31 only, is completely excluded by the radio limits.

One possible caveat to the conclusions above is the uncertainties in the fitting cross section values due to the uncertainties in J-factor values for both MW and M31. These uncertainties are written out in [\cite{2021PhRvD.103b3027K}, Table II, columns 7--9] and are related mainly to a possible asphericity of DM halos of both galaxies. We can see from those columns, that the cross section may deviate by up to 2--3 times from the average fitting value in the last two columns. This deviation may stretch the green rectangle in Fig. \ref{fig:sv} over the vertical direction. The shift of the upper rectangle border does not matter, since the upper cross section values are absolutely excluded. Only the bottom border has relevance. The bottom border is defined by the first row of [\cite{2021PhRvD.103b3027K}, Table II]. This row shows, that J-factor uncertainties may provide the maximal cross section decrease with respect to the average value by $1.52 \times 1.32 \times 1.38 \approx 2.8$ times. However, at the same time, there is another systematical uncertainty of observational origin. \cite{2021PhRvD.103b3027K} analyzed individually the whole OH (SH in their terminology) and its northern/southern halves (SHN/SHS), and obtained the fitting cross sections for all three cases. The cross section values for the whole halo and the southern half match quite well. But the northern half requires the cross section value by $\approx 2$ times larger. This discrepancy can be attributed to significant observational systematics due to an overall weakness of the purported OH emission. Here I would like to note, that there is a mistake in the text of \cite{2021PhRvD.103b3027K}, where the cross section ratios for the different halo parts are described. This mistake was confirmed by the author of \cite{2021PhRvD.103b3027K} in private communication. Thus, the text must state "SH/SHS $\approx 1$" and "SH/SHN $\approx 0.5$" instead of "SH/SHS = 1.8" and "SH/SHN = 1.0". Table II there provides the cross section values for the whole halo (SH), i.e. the minimal values with respect to this observational systematics, since SHN requires $\approx$ 2 times higher cross section. Therefore, this observational systematics is able to counteract the cross section decrease by 2.8 times, which may be provided by J-factor uncertainties described above. Summarizing, we would need in extremely fine-tuned circumstances in order to push the bottom border of the green rectangle significantly lower in Fig. \ref{fig:sv}. These circumstances include: an extreme boost factor configuration, high and tuned asphericity of DM halos of both galaxies, favorably tuned observational systematics and other. A random coincidence of all these factors is unrealistic. Thus, the caveat described in this paragraph does not seem to be able to change noticeably our conclusions about certain tension with the radio constraints.

Summarizing a somewhat qualitative discussion in this section, we can state the following. The cuspy M31 DM density profiles like Einasto MAX can not fit the outer gamma-ray halo without violation of the radio constraints. But rather cored profiles may accommodate the requirements from both radio and gamma-ray bands, if the latter would require the fitting cross section values comparable to those for the cuspy profiles. Such cored profiles are possible due to, e.g., sophisticated effects of long-term washing out of the initial cusp \cite{2021ApJ...919...86B} and are not excluded observationally. However, the cored profiles would likely require a significant or even dominant gamma-ray signal contribution from MW along the OH line of sight and quite high boost factor (at least in MW) for the OH fit. This follows from [\cite{2021PhRvD.103b3027K}, Table II] and the fact, that high cross sections $\langle \sigma v \rangle \gtrsim 10^{-25}$ cm$^3$/s are robustly excluded for any scenarios in Fig. \ref{fig:sv}. And, finally, just to clarify, the discussion here does not make any inferences on the cuspy/cored property of MW DM density profile in the context of OH fit.

\section{\label{sec:con}Conclusions and discussion}

The main goal of this work was to model theoretically the interesting effect of the asymmetry of ICS emission from a population of relativistic $e^\pm$ in M31 halo. This effect was pointed out for the first time in \cite{2020arXiv201104689B}. The asymmetry arises, because the ISRF of a disk galaxy is anisotropic, and we view M31 under certain inclination angle. The asymmetry appears as the difference in ICS emission intensity between the points, which are located symmetrically with respect to the major galactic axis.

ICS emission intensity maps were computed by GALPROP code (v57). As the source of $e^\pm$ population, the thermal WIMPs with $m_x = 60$ GeV annihilating to $b\bar{b}$ were set. This WIMP may be responsible for the gamma-ray emission from M31 OH according to \cite{2021PhRvD.103b3027K}. This $e^\pm$ population was chosen as a trial example to study the effect in general and, at the same time, in order to build specific observational predictions for DM interpretation of the OH emission. Some alternative $e^\pm$ populations from DM were studied too.

The conducted computation confirmed the asymmetry effect presence at a significant level. Table \ref{tab:as} displays the main results of calculations: the asymmetry dependence on the energy and model parameters in the IH and OH ROIs. We can emphasize the following key properties of the asymmetry: 1) it appears at the energies $E_\gamma \gtrsim 0.1$ MeV; 2) has the opposite sign in the IH and OH ROIs; 3) by absolute value it is comparable in the IH and OH ROIs; 4) has a moderate sensitivity to the galactic and $e^\pm$ population model parameters; 5) ranges at most from 0 to $\approx$ 30\% by absolute value in the IH and OH ROIs; 6) addition of the prompt emission from DM annihilation vanishes the asymmetry at $E_\gamma \gtrsim 1$ GeV. 

A potential observational detection of the asymmetry would provide valuable inferences for the emission mechanism understanding. We can outline the following semi-qualitative classification in this regard:
\begin{enumerate}
	\item If a diffuse gamma-ray emission is absolutely symmetric at all energies with respect to the major axis of M31, this implies likely a purely hadronic emission mechanism.
	\item A mild emission asymmetry at the level of few \% implies likely a mixed leptonic/hadronic emission scenario.
	\item A significant asymmetry at the level $\gtrsim$ 10\% implies a dominance of the leptonic sources; which may include $e^\pm$ from CRs, DM and even probably MSPs.
	\item The asymmetry in the case of DM $e^\pm$ appears over a significantly narrower energy range than in the case of CR $e^\pm$.
\end{enumerate}
This scheme is outlined in the context of considered ROIs, indeed, and does not necessarily need to be applied to both of them simultaneously. Different parts of the galactic halo may have different emission mechanisms, and the latter can be tested through the asymmetry measurements regionally. Overall, the logic here indeed applies to only a large-scale diffuse emission and does not apply to an emission from discrete sources. And, of course, the above scheme is majorly idealistic in the sense, that it ignores possible nuisance asymmetries due to, e.g., inhomogeneities of the galactic medium. Considering specifically DM interpretation of the outer gamma-ray halo, this emission component is expected to manifest the following distinct feature: $\approx$ 10\% asymmetry in the energy range $\sim$(1--100) MeV in both the IH and OH ROIs (see the third row of Table \ref{tab:as}). Thus, potential asymmetry measurements may provide a valuable observational tool for diagnosing the emission generation mechanism. Particularly, such tool may confirm DM origin of the emission and, thus, unravel DM nature eventually. However, from an observational point of view, such measurements are very challenging and require very high sensitivity; since M31 is rather faint object on the gamma-ray sky. It is generally hard to separate various emission components inside M31. And it would be especially difficult to detect the asymmetry of the potential DM signal, since its ICS component has the intensity much lower than both the Galactic foreground and isotropic background, as illustrated in Fig. \ref{fig:sp}. And hadronic emissions inside M31 may create additional nuisance background too. Summarizing, the revealed ICS emission asymmetry effect is rather theoretical so far, since its observational detection may require virtually unlimited sensitivity. However, if, for example, the brightest (central) region of M31 is shining through ICS, the asymmetry detection there may be achievable in the near future. In general, it is a very sophisticated task to model the quantitative observational predictions for any specific gamma-ray telescope, this is beyond the scope of this work.

Meanwhile, as was already mentioned in Sec. \ref{sec:dm}, \cite{2021PhRvD.103b3027K} reported the observed intensity difference between the northern and southern halves of OH by $\approx$ 2 times. One may naturally ask whether this observed asymmetry can be explained by the ICS emission asymmetry being discussed. Considering DM as the source of $e^\pm$, the answer is absolutely no, since, as was noted above, the bright prompt DM emission component vanishes the asymmetry in the energy range $E_\gamma \geqslant 1$ GeV employed in \cite{2021PhRvD.103b3027K}.

The ICS emission asymmetry effect would, in principle, exist in any disk galaxy, which is viewed under the inclination angle different from 0$\degree$ or 90$\degree$. Examples of such nearby galaxies include Large Magellanic Cloud (LMC), M81, M33 and others. However, all of them except LMC are farther than M31. Hence, it would be even more difficult to study their gamma-ray emission details.

A potential direction for further development of this topic is to conduct the same asymmetry modeling for the case of CR $e^\pm$ population and compare the results with DM case. The results may differ due to the differences in both the $e^\pm$ source spatial distribution and energy spectrum at injection.

Other sections of the paper discussed other aspects, which are directly relevant for DM interpretation of the outer gamma-ray halo. One aspect is the role of the secondary, i.e. ICS and bremsstrahlung, emissions due to DM annihilation in both MW and M31. It was proved that these secondary components can be neglected in the fit of OH in the relevant energy range $E_\gamma \gtrsim 1$ GeV for the case of nearly thermal annihilation cross section. However, at lower energies or/and in the case of higher cross sections, one may need to take into account some of the secondary components. The second aspect is the relation between WIMP radio constraints derived recently in \cite{2022PhRvD.106b3023E} and the outer gamma-ray halo fit. These constraints imply certain restrictions on the possibility of the fit (from Fig. \ref{fig:sv}): a significant contribution from MW to the gamma-ray signal along M31 OH line of sight is inevitably necessary, as well as quite high boost factor due to substructures (at least in MW DM halo). Regarding M31, only rather cored DM density profiles are suitable. Overall, we may conclude, that DM explanation of M31 OH is still possible in a view of radio constraints. However, this explanation does not look all that natural, since it requires certain fine tuning of the model ingredients. Particularly, a bit optimistic boost factor values are needed. Also, WIMP constraints from observations of other objects, particularly dwarf MW satellites, may provide another independent restrictions. Thus, it is probably premature now to consider M31 outer gamma-ray halo phenomenon as definite DM manifestation. More observational data on M31 is needed in order to unravel large systematical uncertainties and understand the emission mechanisms better.    

It is indeed difficult to provide full coverage of such a complicated topic in the frame of one single paper. The modeling here was tied mainly to the quite standard thermal WIMPs and the diffusion of their annihilation products. However, more exotic scenarios may be relevant. E.g., the so-called leptophilic DM \cite{2021PhRvD.104g5016G}, which produces a lot of $e^\pm$, which in turn would produce a bright ICS emission. Regarding the diffusion model, one may consider a larger size of the diffusion zone (e.g., \cite{2021PhRvD.104l3016D}).

\begin{acknowledgments}
I dedicate this paper to the kind memory of \textbf{Nikolay Topchiev} and \textbf{Arkadii Galper}, who recently passed away. They had been working in the field of experimental high-energy astrophysics for nearly 50 and 70 years, respectively.	They will always be remembered as enthusiastic scientists and great supervisors, fellows and friends. 
	
I greatly appreciate the following software, which is very useful: Wolfram Mathematica; "Aladin sky atlas" developed at CDS, Strasbourg Observatory, France \cite{2000A&AS..143...33B}; NumPy \cite{numpy}; Healpy \cite{healpy}; WebPlotDigitizer \cite{WPD}.
 
I am very grateful to Alexei Alexeev for providing access to high-performance computing cluster, Mikhail Zavertyaev for providing another server and also to Mikhail Razumeyko.
\end{acknowledgments}

\bibliography{../../../universal}

\begin{thebibliography}{53}%
\makeatletter
\providecommand \@ifxundefined [1]{%
 \@ifx{#1\undefined}
}%
\providecommand \@ifnum [1]{%
 \ifnum #1\expandafter \@firstoftwo
 \else \expandafter \@secondoftwo
 \fi
}%
\providecommand \@ifx [1]{%
 \ifx #1\expandafter \@firstoftwo
 \else \expandafter \@secondoftwo
 \fi
}%
\providecommand \natexlab [1]{#1}%
\providecommand \enquote  [1]{``#1''}%
\providecommand \bibnamefont  [1]{#1}%
\providecommand \bibfnamefont [1]{#1}%
\providecommand \citenamefont [1]{#1}%
\providecommand \href@noop [0]{\@secondoftwo}%
\providecommand \href [0]{\begingroup \@sanitize@url \@href}%
\providecommand \@href[1]{\@@startlink{#1}\@@href}%
\providecommand \@@href[1]{\endgroup#1\@@endlink}%
\providecommand \@sanitize@url [0]{\catcode `\\12\catcode `\$12\catcode
  `\&12\catcode `\#12\catcode `\^12\catcode `\_12\catcode `\%12\relax}%
\providecommand \@@startlink[1]{}%
\providecommand \@@endlink[0]{}%
\providecommand \url  [0]{\begingroup\@sanitize@url \@url }%
\providecommand \@url [1]{\endgroup\@href {#1}{\urlprefix }}%
\providecommand \urlprefix  [0]{URL }%
\providecommand \Eprint [0]{\href }%
\providecommand \doibase [0]{https://doi.org/}%
\providecommand \selectlanguage [0]{\@gobble}%
\providecommand \bibinfo  [0]{\@secondoftwo}%
\providecommand \bibfield  [0]{\@secondoftwo}%
\providecommand \translation [1]{[#1]}%
\providecommand \BibitemOpen [0]{}%
\providecommand \bibitemStop [0]{}%
\providecommand \bibitemNoStop [0]{.\EOS\space}%
\providecommand \EOS [0]{\spacefactor3000\relax}%
\providecommand \BibitemShut  [1]{\csname bibitem#1\endcsname}%
\let\auto@bib@innerbib\@empty
\bibitem [{\citenamefont {{Abdo}}\ \emph {et~al.}(2010)\citenamefont {{Abdo}},
  \citenamefont {{Ackermann}}, \citenamefont {{Ajello}}, \citenamefont
  {{Allafort}}, \citenamefont {{Atwood}}, \citenamefont {{Baldini}},
  \citenamefont {{Ballet}}, \citenamefont {{Barbiellini}}, \citenamefont
  {{Bastieri}}, \citenamefont {{Bechtol}}, \citenamefont {{Bellazzini}},
  \citenamefont {{Berenji}}, \citenamefont {{Blandford}}, \citenamefont
  {{Bloom}}, \citenamefont {{Bonamente}} \emph {et~al.}}]{2010A&A...523L...2A}%
  \BibitemOpen
  \bibfield  {author} {\bibinfo {author} {\bibfnamefont {A.~A.}\ \bibnamefont
  {{Abdo}}}, \bibinfo {author} {\bibfnamefont {M.}~\bibnamefont {{Ackermann}}},
  \bibinfo {author} {\bibfnamefont {M.}~\bibnamefont {{Ajello}}}, \bibinfo
  {author} {\bibfnamefont {A.}~\bibnamefont {{Allafort}}}, \bibinfo {author}
  {\bibfnamefont {W.~B.}\ \bibnamefont {{Atwood}}}, \bibinfo {author}
  {\bibfnamefont {L.}~\bibnamefont {{Baldini}}}, \bibinfo {author}
  {\bibfnamefont {J.}~\bibnamefont {{Ballet}}}, \bibinfo {author}
  {\bibfnamefont {G.}~\bibnamefont {{Barbiellini}}}, \bibinfo {author}
  {\bibfnamefont {D.}~\bibnamefont {{Bastieri}}}, \bibinfo {author}
  {\bibfnamefont {K.}~\bibnamefont {{Bechtol}}}, \bibinfo {author}
  {\bibfnamefont {R.}~\bibnamefont {{Bellazzini}}}, \bibinfo {author}
  {\bibfnamefont {B.}~\bibnamefont {{Berenji}}}, \bibinfo {author}
  {\bibfnamefont {R.~D.}\ \bibnamefont {{Blandford}}}, \bibinfo {author}
  {\bibfnamefont {E.~D.}\ \bibnamefont {{Bloom}}}, \bibinfo {author}
  {\bibfnamefont {E.}~\bibnamefont {{Bonamente}}}, \emph {et~al.},\ }\href
  {https://doi.org/10.1051/0004-6361/201015759} {\bibfield  {journal} {\bibinfo
   {journal} {Astron. Astrophys.}\ }\textbf {\bibinfo {volume} {523}},\
  \bibinfo {eid} {L2} (\bibinfo {year} {2010})},\ \Eprint
  {https://arxiv.org/abs/1012.1952} {arXiv:1012.1952 [astro-ph.HE]}
  \BibitemShut {NoStop}%
\bibitem [{\citenamefont {{Ackermann}}\ \emph {et~al.}(2017)\citenamefont
  {{Ackermann}}, \citenamefont {{Ajello}}, \citenamefont {{Albert}},
  \citenamefont {{Baldini}}, \citenamefont {{Ballet}}, \citenamefont
  {{Barbiellini}}, \citenamefont {{Bastieri}}, \citenamefont {{Bellazzini}},
  \citenamefont {{Bissaldi}}, \citenamefont {{Bloom}}, \citenamefont
  {{Bonino}}, \citenamefont {{Bottacini}}, \citenamefont {{Brandt}},
  \citenamefont {{Bregeon}}, \citenamefont {{Bruel}} \emph
  {et~al.}}]{2017ApJ...836..208A}%
  \BibitemOpen
  \bibfield  {author} {\bibinfo {author} {\bibfnamefont {M.}~\bibnamefont
  {{Ackermann}}}, \bibinfo {author} {\bibfnamefont {M.}~\bibnamefont
  {{Ajello}}}, \bibinfo {author} {\bibfnamefont {A.}~\bibnamefont {{Albert}}},
  \bibinfo {author} {\bibfnamefont {L.}~\bibnamefont {{Baldini}}}, \bibinfo
  {author} {\bibfnamefont {J.}~\bibnamefont {{Ballet}}}, \bibinfo {author}
  {\bibfnamefont {G.}~\bibnamefont {{Barbiellini}}}, \bibinfo {author}
  {\bibfnamefont {D.}~\bibnamefont {{Bastieri}}}, \bibinfo {author}
  {\bibfnamefont {R.}~\bibnamefont {{Bellazzini}}}, \bibinfo {author}
  {\bibfnamefont {E.}~\bibnamefont {{Bissaldi}}}, \bibinfo {author}
  {\bibfnamefont {E.~D.}\ \bibnamefont {{Bloom}}}, \bibinfo {author}
  {\bibfnamefont {R.}~\bibnamefont {{Bonino}}}, \bibinfo {author}
  {\bibfnamefont {E.}~\bibnamefont {{Bottacini}}}, \bibinfo {author}
  {\bibfnamefont {T.~J.}\ \bibnamefont {{Brandt}}}, \bibinfo {author}
  {\bibfnamefont {J.}~\bibnamefont {{Bregeon}}}, \bibinfo {author}
  {\bibfnamefont {P.}~\bibnamefont {{Bruel}}}, \emph {et~al.},\ }\href
  {https://doi.org/10.3847/1538-4357/aa5c3d} {\bibfield  {journal} {\bibinfo
  {journal} {Astrophys. J.}\ }\textbf {\bibinfo {volume} {836}},\ \bibinfo
  {eid} {208} (\bibinfo {year} {2017})},\ \Eprint
  {https://arxiv.org/abs/1702.08602} {arXiv:1702.08602 [astro-ph.HE]}
  \BibitemShut {NoStop}%
\bibitem [{\citenamefont {{Armand}}\ and\ \citenamefont
  {{Calore}}(2021)}]{2021PhRvD.103h3023A}%
  \BibitemOpen
  \bibfield  {author} {\bibinfo {author} {\bibfnamefont {C.}~\bibnamefont
  {{Armand}}}\ and\ \bibinfo {author} {\bibfnamefont {F.}~\bibnamefont
  {{Calore}}},\ }\href {https://doi.org/10.1103/PhysRevD.103.083023} {\bibfield
   {journal} {\bibinfo  {journal} {Phys. Rev. D}\ }\textbf {\bibinfo {volume}
  {103}},\ \bibinfo {eid} {083023} (\bibinfo {year} {2021})},\ \Eprint
  {https://arxiv.org/abs/2102.06447} {arXiv:2102.06447 [astro-ph.HE]}
  \BibitemShut {NoStop}%
\bibitem [{\citenamefont {{Karwin}}\ \emph {et~al.}(2019)\citenamefont
  {{Karwin}}, \citenamefont {{Murgia}}, \citenamefont {{Campbell}},\ and\
  \citenamefont {{Moskalenko}}}]{2019ApJ...880...95K}%
  \BibitemOpen
  \bibfield  {author} {\bibinfo {author} {\bibfnamefont {C.~M.}\ \bibnamefont
  {{Karwin}}}, \bibinfo {author} {\bibfnamefont {S.}~\bibnamefont {{Murgia}}},
  \bibinfo {author} {\bibfnamefont {S.}~\bibnamefont {{Campbell}}},\ and\
  \bibinfo {author} {\bibfnamefont {I.~V.}\ \bibnamefont {{Moskalenko}}},\
  }\href {https://doi.org/10.3847/1538-4357/ab2880} {\bibfield  {journal}
  {\bibinfo  {journal} {Astrophys. J.}\ }\textbf {\bibinfo {volume} {880}},\
  \bibinfo {eid} {95} (\bibinfo {year} {2019})},\ \Eprint
  {https://arxiv.org/abs/1812.02958} {arXiv:1812.02958 [astro-ph.HE]}
  \BibitemShut {NoStop}%
\bibitem [{\citenamefont {{Xing}}\ \emph {et~al.}(2023)\citenamefont {{Xing}},
  \citenamefont {{Wang}}, \citenamefont {{Zheng}},\ and\ \citenamefont
  {{Li}}}]{2023ApJ...945L..22X}%
  \BibitemOpen
  \bibfield  {author} {\bibinfo {author} {\bibfnamefont {Y.}~\bibnamefont
  {{Xing}}}, \bibinfo {author} {\bibfnamefont {Z.}~\bibnamefont {{Wang}}},
  \bibinfo {author} {\bibfnamefont {D.}~\bibnamefont {{Zheng}}},\ and\ \bibinfo
  {author} {\bibfnamefont {J.}~\bibnamefont {{Li}}},\ }\href
  {https://doi.org/10.3847/2041-8213/acbf4f} {\bibfield  {journal} {\bibinfo
  {journal} {ApJL}\ }\textbf {\bibinfo {volume} {945}},\ \bibinfo {eid} {L22}
  (\bibinfo {year} {2023})},\ \Eprint {https://arxiv.org/abs/2301.06743}
  {arXiv:2301.06743 [astro-ph.HE]} \BibitemShut {NoStop}%
\bibitem [{\citenamefont {{Zimmer}}\ \emph {et~al.}(2022)\citenamefont
  {{Zimmer}}, \citenamefont {{Macias}}, \citenamefont {{Ando}}, \citenamefont
  {{Crocker}},\ and\ \citenamefont {{Horiuchi}}}]{2022MNRAS.516.4469Z}%
  \BibitemOpen
  \bibfield  {author} {\bibinfo {author} {\bibfnamefont {F.}~\bibnamefont
  {{Zimmer}}}, \bibinfo {author} {\bibfnamefont {O.}~\bibnamefont {{Macias}}},
  \bibinfo {author} {\bibfnamefont {S.}~\bibnamefont {{Ando}}}, \bibinfo
  {author} {\bibfnamefont {R.~M.}\ \bibnamefont {{Crocker}}},\ and\ \bibinfo
  {author} {\bibfnamefont {S.}~\bibnamefont {{Horiuchi}}},\ }\href
  {https://doi.org/10.1093/mnras/stac2464} {\bibfield  {journal} {\bibinfo
  {journal} {Mon. Not. R. Astron. Soc.}\ }\textbf {\bibinfo {volume} {516}},\
  \bibinfo {pages} {4469} (\bibinfo {year} {2022})},\ \Eprint
  {https://arxiv.org/abs/2204.00636} {arXiv:2204.00636 [astro-ph.HE]}
  \BibitemShut {NoStop}%
\bibitem [{\citenamefont {{McDaniel}}\ \emph {et~al.}(2019)\citenamefont
  {{McDaniel}}, \citenamefont {{Jeltema}},\ and\ \citenamefont
  {{Profumo}}}]{2019PhRvD.100b3014M}%
  \BibitemOpen
  \bibfield  {author} {\bibinfo {author} {\bibfnamefont {A.}~\bibnamefont
  {{McDaniel}}}, \bibinfo {author} {\bibfnamefont {T.}~\bibnamefont
  {{Jeltema}}},\ and\ \bibinfo {author} {\bibfnamefont {S.}~\bibnamefont
  {{Profumo}}},\ }\href {https://doi.org/10.1103/PhysRevD.100.023014}
  {\bibfield  {journal} {\bibinfo  {journal} {Phys. Rev. D}\ }\textbf {\bibinfo
  {volume} {100}},\ \bibinfo {eid} {023014} (\bibinfo {year} {2019})},\ \Eprint
  {https://arxiv.org/abs/1903.06833} {arXiv:1903.06833 [astro-ph.HE]}
  \BibitemShut {NoStop}%
\bibitem [{\citenamefont {{Do}}\ \emph {et~al.}(2021)\citenamefont {{Do}},
  \citenamefont {{Duong}}, \citenamefont {{McDaniel}}, \citenamefont
  {{O'Connor}}, \citenamefont {{Profumo}}, \citenamefont {{Rafael}},
  \citenamefont {{Sweeney}},\ and\ \citenamefont
  {{Vera}}}]{2021PhRvD.104l3016D}%
  \BibitemOpen
  \bibfield  {author} {\bibinfo {author} {\bibfnamefont {A.}~\bibnamefont
  {{Do}}}, \bibinfo {author} {\bibfnamefont {M.}~\bibnamefont {{Duong}}},
  \bibinfo {author} {\bibfnamefont {A.}~\bibnamefont {{McDaniel}}}, \bibinfo
  {author} {\bibfnamefont {C.}~\bibnamefont {{O'Connor}}}, \bibinfo {author}
  {\bibfnamefont {S.}~\bibnamefont {{Profumo}}}, \bibinfo {author}
  {\bibfnamefont {J.}~\bibnamefont {{Rafael}}}, \bibinfo {author}
  {\bibfnamefont {C.}~\bibnamefont {{Sweeney}}},\ and\ \bibinfo {author}
  {\bibfnamefont {W.}~\bibnamefont {{Vera}}},\ }\href
  {https://doi.org/10.1103/PhysRevD.104.123016} {\bibfield  {journal} {\bibinfo
   {journal} {Phys. Rev. D}\ }\textbf {\bibinfo {volume} {104}},\ \bibinfo
  {eid} {123016} (\bibinfo {year} {2021})},\ \Eprint
  {https://arxiv.org/abs/2012.14507} {arXiv:2012.14507 [astro-ph.HE]}
  \BibitemShut {NoStop}%
\bibitem [{\citenamefont {{Karwin}}\ \emph {et~al.}(2021)\citenamefont
  {{Karwin}}, \citenamefont {{Murgia}}, \citenamefont {{Moskalenko}},
  \citenamefont {{Fillingham}}, \citenamefont {{Burns}},\ and\ \citenamefont
  {{Fieg}}}]{2021PhRvD.103b3027K}%
  \BibitemOpen
  \bibfield  {author} {\bibinfo {author} {\bibfnamefont {C.~M.}\ \bibnamefont
  {{Karwin}}}, \bibinfo {author} {\bibfnamefont {S.}~\bibnamefont {{Murgia}}},
  \bibinfo {author} {\bibfnamefont {I.~V.}\ \bibnamefont {{Moskalenko}}},
  \bibinfo {author} {\bibfnamefont {S.~P.}\ \bibnamefont {{Fillingham}}},
  \bibinfo {author} {\bibfnamefont {A.-K.}\ \bibnamefont {{Burns}}},\ and\
  \bibinfo {author} {\bibfnamefont {M.}~\bibnamefont {{Fieg}}},\ }\href
  {https://doi.org/10.1103/PhysRevD.103.023027} {\bibfield  {journal} {\bibinfo
   {journal} {Phys. Rev. D}\ }\textbf {\bibinfo {volume} {103}},\ \bibinfo
  {eid} {023027} (\bibinfo {year} {2021})},\ \Eprint
  {https://arxiv.org/abs/2010.08563} {arXiv:2010.08563 [astro-ph.HE]}
  \BibitemShut {NoStop}%
\bibitem [{\citenamefont {{Li}}\ \emph {et~al.}(2016)\citenamefont {{Li}},
  \citenamefont {{Huang}}, \citenamefont {{Yuan}},\ and\ \citenamefont
  {{Xu}}}]{2016JCAP...12..028L}%
  \BibitemOpen
  \bibfield  {author} {\bibinfo {author} {\bibfnamefont {Z.}~\bibnamefont
  {{Li}}}, \bibinfo {author} {\bibfnamefont {X.}~\bibnamefont {{Huang}}},
  \bibinfo {author} {\bibfnamefont {Q.}~\bibnamefont {{Yuan}}},\ and\ \bibinfo
  {author} {\bibfnamefont {Y.}~\bibnamefont {{Xu}}},\ }\href
  {https://doi.org/10.1088/1475-7516/2016/12/028} {\bibfield  {journal}
  {\bibinfo  {journal} {{J. Cosmol. Astropart. Phys.}}\ }\textbf {\bibinfo
  {volume} {2016}},\ \bibinfo {eid} {028} (\bibinfo {year} {2016})},\ \Eprint
  {https://arxiv.org/abs/1312.7609} {arXiv:1312.7609 [astro-ph.CO]}
  \BibitemShut {NoStop}%
\bibitem [{\citenamefont {{McDaniel}}\ \emph {et~al.}(2018)\citenamefont
  {{McDaniel}}, \citenamefont {{Jeltema}},\ and\ \citenamefont
  {{Profumo}}}]{2018PhRvD..97j3021M}%
  \BibitemOpen
  \bibfield  {author} {\bibinfo {author} {\bibfnamefont {A.}~\bibnamefont
  {{McDaniel}}}, \bibinfo {author} {\bibfnamefont {T.}~\bibnamefont
  {{Jeltema}}},\ and\ \bibinfo {author} {\bibfnamefont {S.}~\bibnamefont
  {{Profumo}}},\ }\href {https://doi.org/10.1103/PhysRevD.97.103021} {\bibfield
   {journal} {\bibinfo  {journal} {Phys. Rev. D}\ }\textbf {\bibinfo {volume}
  {97}},\ \bibinfo {eid} {103021} (\bibinfo {year} {2018})},\ \Eprint
  {https://arxiv.org/abs/1802.05258} {arXiv:1802.05258 [astro-ph.HE]}
  \BibitemShut {NoStop}%
\bibitem [{\citenamefont {{Di Mauro}}\ \emph {et~al.}(2019)\citenamefont {{Di
  Mauro}}, \citenamefont {{Hou}}, \citenamefont {{Eckner}}, \citenamefont
  {{Zaharijas}},\ and\ \citenamefont {{Charles}}}]{2019PhRvD..99l3027D}%
  \BibitemOpen
  \bibfield  {author} {\bibinfo {author} {\bibfnamefont {M.}~\bibnamefont {{Di
  Mauro}}}, \bibinfo {author} {\bibfnamefont {X.}~\bibnamefont {{Hou}}},
  \bibinfo {author} {\bibfnamefont {C.}~\bibnamefont {{Eckner}}}, \bibinfo
  {author} {\bibfnamefont {G.}~\bibnamefont {{Zaharijas}}},\ and\ \bibinfo
  {author} {\bibfnamefont {E.}~\bibnamefont {{Charles}}},\ }\href
  {https://doi.org/10.1103/PhysRevD.99.123027} {\bibfield  {journal} {\bibinfo
  {journal} {Phys. Rev. D}\ }\textbf {\bibinfo {volume} {99}},\ \bibinfo {eid}
  {123027} (\bibinfo {year} {2019})},\ \Eprint
  {https://arxiv.org/abs/1904.10977} {arXiv:1904.10977 [astro-ph.HE]}
  \BibitemShut {NoStop}%
\bibitem [{\citenamefont {{Belotsky}}\ \emph {et~al.}(2020)\citenamefont
  {{Belotsky}}, \citenamefont {{Shlepkina}},\ and\ \citenamefont
  {{Soloviev}}}]{2020arXiv201104689B}%
  \BibitemOpen
  \bibfield  {author} {\bibinfo {author} {\bibfnamefont {K.~M.}\ \bibnamefont
  {{Belotsky}}}, \bibinfo {author} {\bibfnamefont {E.~S.}\ \bibnamefont
  {{Shlepkina}}},\ and\ \bibinfo {author} {\bibfnamefont {M.~L.}\ \bibnamefont
  {{Soloviev}}},\ }\href {https://doi.org/10.48550/arXiv.2011.04689} {\bibfield
   {journal} {\bibinfo  {journal} {arXiv}\ ,\ \bibinfo {eid}
  {arXiv:2011.04689}} (\bibinfo {year} {2020})},\ \Eprint
  {https://arxiv.org/abs/2011.04689} {arXiv:2011.04689 [astro-ph.HE]}
  \BibitemShut {NoStop}%
\bibitem [{GP()}]{GP}%
  \BibitemOpen
  \href@noop {} {}\bibinfo {howpublished}
  {\url{http://galprop.stanford.edu/}}\BibitemShut {NoStop}%
\bibitem [{\citenamefont {{Egorov}}\ and\ \citenamefont
  {{Pierpaoli}}(2013)}]{2013PhRvD..88b3504E}%
  \BibitemOpen
  \bibfield  {author} {\bibinfo {author} {\bibfnamefont {A.~E.}\ \bibnamefont
  {{Egorov}}}\ and\ \bibinfo {author} {\bibfnamefont {E.}~\bibnamefont
  {{Pierpaoli}}},\ }\href {https://doi.org/10.1103/PhysRevD.88.023504}
  {\bibfield  {journal} {\bibinfo  {journal} {Phys. Rev. D}\ }\textbf {\bibinfo
  {volume} {88}},\ \bibinfo {eid} {023504} (\bibinfo {year} {2013})},\ \Eprint
  {https://arxiv.org/abs/1304.0517} {arXiv:1304.0517 [astro-ph.CO]}
  \BibitemShut {NoStop}%
\bibitem [{\citenamefont {{Egorov}}(2022{\natexlab{a}})}]{2022PhRvD.106b3023E}%
  \BibitemOpen
  \bibfield  {author} {\bibinfo {author} {\bibfnamefont {A.~E.}\ \bibnamefont
  {{Egorov}}},\ }\href {https://doi.org/10.1103/PhysRevD.106.023023} {\bibfield
   {journal} {\bibinfo  {journal} {Phys. Rev. D}\ }\textbf {\bibinfo {volume}
  {106}},\ \bibinfo {eid} {023023} (\bibinfo {year} {2022}{\natexlab{a}})},\
  \Eprint {https://arxiv.org/abs/2205.01033} {arXiv:2205.01033 [astro-ph.CO]}
  \BibitemShut {NoStop}%
\bibitem [{\citenamefont {{Egorov}}(2022{\natexlab{b}})}]{2022arXiv220814186E}%
  \BibitemOpen
  \bibfield  {author} {\bibinfo {author} {\bibfnamefont {A.~E.}\ \bibnamefont
  {{Egorov}}},\ }\href {https://doi.org/10.48550/arXiv.2208.14186} {\bibfield
  {journal} {\bibinfo  {journal} {arXiv}\ ,\ \bibinfo {eid} {arXiv:2208.14186}}
  (\bibinfo {year} {2022}{\natexlab{b}})},\ \Eprint
  {https://arxiv.org/abs/2208.14186} {arXiv:2208.14186 [hep-ph]} \BibitemShut
  {NoStop}%
\bibitem [{\citenamefont {Berestetskii}\ \emph {et~al.}(1989)\citenamefont
  {Berestetskii}, \citenamefont {Lifshitz},\ and\ \citenamefont
  {Pitaevskii}}]{LL4}%
  \BibitemOpen
  \bibfield  {author} {\bibinfo {author} {\bibfnamefont {V.}~\bibnamefont
  {Berestetskii}}, \bibinfo {author} {\bibfnamefont {E.}~\bibnamefont
  {Lifshitz}},\ and\ \bibinfo {author} {\bibfnamefont {L.}~\bibnamefont
  {Pitaevskii}},\ }\href@noop {} {\emph {\bibinfo {title} {Quantum
  electrodynamics (vol. IV)}}}\ (\bibinfo  {publisher} {Nauka, Moscow (in
  Russian)},\ \bibinfo {year} {1989})\BibitemShut {NoStop}%
\bibitem [{\citenamefont {Ginzburg}(1979)}]{Ginzburg}%
  \BibitemOpen
  \bibfield  {author} {\bibinfo {author} {\bibfnamefont {V.}~\bibnamefont
  {Ginzburg}},\ }\href@noop {} {\emph {\bibinfo {title} {Theoretical physics
  and astrophysics}}}\ (\bibinfo  {publisher} {Pergamon press},\ \bibinfo
  {year} {1979})\BibitemShut {NoStop}%
\bibitem [{sv()}]{sv}%
  \BibitemOpen
  \href@noop {} {}\bibinfo {howpublished}
  {\url{https://member.ipmu.jp/satoshi.shirai/DM2020/}}\BibitemShut {NoStop}%
\bibitem [{\citenamefont {{Tamm}}\ \emph {et~al.}(2012)\citenamefont {{Tamm}},
  \citenamefont {{Tempel}}, \citenamefont {{Tenjes}}, \citenamefont
  {{Tihhonova}},\ and\ \citenamefont {{Tuvikene}}}]{2012A&A...546A...4T}%
  \BibitemOpen
  \bibfield  {author} {\bibinfo {author} {\bibfnamefont {A.}~\bibnamefont
  {{Tamm}}}, \bibinfo {author} {\bibfnamefont {E.}~\bibnamefont {{Tempel}}},
  \bibinfo {author} {\bibfnamefont {P.}~\bibnamefont {{Tenjes}}}, \bibinfo
  {author} {\bibfnamefont {O.}~\bibnamefont {{Tihhonova}}},\ and\ \bibinfo
  {author} {\bibfnamefont {T.}~\bibnamefont {{Tuvikene}}},\ }\href
  {https://doi.org/10.1051/0004-6361/201220065} {\bibfield  {journal} {\bibinfo
   {journal} {Astron. Astrophys.}\ }\textbf {\bibinfo {volume} {546}},\
  \bibinfo {eid} {A4} (\bibinfo {year} {2012})},\ \Eprint
  {https://arxiv.org/abs/1208.5712} {arXiv:1208.5712 [astro-ph.CO]}
  \BibitemShut {NoStop}%
\bibitem [{\citenamefont {{Kamionkowski}}\ \emph {et~al.}(2010)\citenamefont
  {{Kamionkowski}}, \citenamefont {{Koushiappas}},\ and\ \citenamefont
  {{Kuhlen}}}]{2010PhRvD..81d3532K}%
  \BibitemOpen
  \bibfield  {author} {\bibinfo {author} {\bibfnamefont {M.}~\bibnamefont
  {{Kamionkowski}}}, \bibinfo {author} {\bibfnamefont {S.~M.}\ \bibnamefont
  {{Koushiappas}}},\ and\ \bibinfo {author} {\bibfnamefont {M.}~\bibnamefont
  {{Kuhlen}}},\ }\href {https://doi.org/10.1103/PhysRevD.81.043532} {\bibfield
  {journal} {\bibinfo  {journal} {Phys. Rev. D}\ }\textbf {\bibinfo {volume}
  {81}},\ \bibinfo {eid} {043532} (\bibinfo {year} {2010})},\ \Eprint
  {https://arxiv.org/abs/1001.3144} {arXiv:1001.3144 [astro-ph.GA]}
  \BibitemShut {NoStop}%
\bibitem [{\citenamefont {{Fornengo}}\ \emph {et~al.}(2012)\citenamefont
  {{Fornengo}}, \citenamefont {{Lineros}}, \citenamefont {{Regis}},\ and\
  \citenamefont {{Taoso}}}]{2012JCAP...01..005F}%
  \BibitemOpen
  \bibfield  {author} {\bibinfo {author} {\bibfnamefont {N.}~\bibnamefont
  {{Fornengo}}}, \bibinfo {author} {\bibfnamefont {R.~A.}\ \bibnamefont
  {{Lineros}}}, \bibinfo {author} {\bibfnamefont {M.}~\bibnamefont {{Regis}}},\
  and\ \bibinfo {author} {\bibfnamefont {M.}~\bibnamefont {{Taoso}}},\ }\href
  {https://doi.org/10.1088/1475-7516/2012/01/005} {\bibfield  {journal}
  {\bibinfo  {journal} {{J. Cosmol. Astropart. Phys.}}\ }\textbf {\bibinfo
  {volume} {01}},\ \bibinfo {eid} {005} (\bibinfo {year} {2012})},\ \Eprint
  {https://arxiv.org/abs/1110.4337} {arXiv:1110.4337 [astro-ph.GA]}
  \BibitemShut {NoStop}%
\bibitem [{\citenamefont {{Porter}}\ \emph {et~al.}(2022)\citenamefont
  {{Porter}}, \citenamefont {{J{\'o}hannesson}},\ and\ \citenamefont
  {{Moskalenko}}}]{2022ApJS..262...30P}%
  \BibitemOpen
  \bibfield  {author} {\bibinfo {author} {\bibfnamefont {T.~A.}\ \bibnamefont
  {{Porter}}}, \bibinfo {author} {\bibfnamefont {G.}~\bibnamefont
  {{J{\'o}hannesson}}},\ and\ \bibinfo {author} {\bibfnamefont {I.~V.}\
  \bibnamefont {{Moskalenko}}},\ }\href
  {https://doi.org/10.3847/1538-4365/ac80f6} {\bibfield  {journal} {\bibinfo
  {journal} {Astrophys. J. Suppl. Ser.}\ }\textbf {\bibinfo {volume} {262}},\
  \bibinfo {eid} {30} (\bibinfo {year} {2022})},\ \Eprint
  {https://arxiv.org/abs/2112.12745} {arXiv:2112.12745 [astro-ph.HE]}
  \BibitemShut {NoStop}%
\bibitem [{git()}]{github}%
  \BibitemOpen
  \href@noop {} {}\bibinfo {howpublished}
  {\url{https://github.com/a-e-egorov/GALPROP_DM}}\BibitemShut {NoStop}%
\bibitem [{PPP()}]{PPPC}%
  \BibitemOpen
  \href@noop {} {}\bibinfo {howpublished}
  {\url{http://www.marcocirelli.net/PPPC4DMID.html}}\BibitemShut {NoStop}%
\bibitem [{\citenamefont {{Cirelli}}\ \emph {et~al.}(2011)\citenamefont
  {{Cirelli}}, \citenamefont {{Corcella}}, \citenamefont {{Hektor}},
  \citenamefont {{H{\"u}tsi}}, \citenamefont {{Kadastik}}, \citenamefont
  {{Panci}}, \citenamefont {{Raidal}}, \citenamefont {{Sala}},\ and\
  \citenamefont {{Strumia}}}]{2011JCAP...03..051C}%
  \BibitemOpen
  \bibfield  {author} {\bibinfo {author} {\bibfnamefont {M.}~\bibnamefont
  {{Cirelli}}}, \bibinfo {author} {\bibfnamefont {G.}~\bibnamefont
  {{Corcella}}}, \bibinfo {author} {\bibfnamefont {A.}~\bibnamefont
  {{Hektor}}}, \bibinfo {author} {\bibfnamefont {G.}~\bibnamefont
  {{H{\"u}tsi}}}, \bibinfo {author} {\bibfnamefont {M.}~\bibnamefont
  {{Kadastik}}}, \bibinfo {author} {\bibfnamefont {P.}~\bibnamefont {{Panci}}},
  \bibinfo {author} {\bibfnamefont {M.}~\bibnamefont {{Raidal}}}, \bibinfo
  {author} {\bibfnamefont {F.}~\bibnamefont {{Sala}}},\ and\ \bibinfo {author}
  {\bibfnamefont {A.}~\bibnamefont {{Strumia}}},\ }\href
  {https://doi.org/10.1088/1475-7516/2011/03/051} {\bibfield  {journal}
  {\bibinfo  {journal} {{J. Cosmol. Astropart. Phys.}}\ }\textbf {\bibinfo
  {volume} {03}},\ \bibinfo {eid} {051} (\bibinfo {year} {2011})},\ \Eprint
  {https://arxiv.org/abs/1012.4515} {arXiv:1012.4515 [hep-ph]} \BibitemShut
  {NoStop}%
\bibitem [{\citenamefont {{Ciafaloni}}\ \emph {et~al.}(2011)\citenamefont
  {{Ciafaloni}}, \citenamefont {{Comelli}}, \citenamefont {{Riotto}},
  \citenamefont {{Sala}}, \citenamefont {{Strumia}},\ and\ \citenamefont
  {{Urbano}}}]{2011JCAP...03..019C}%
  \BibitemOpen
  \bibfield  {author} {\bibinfo {author} {\bibfnamefont {P.}~\bibnamefont
  {{Ciafaloni}}}, \bibinfo {author} {\bibfnamefont {D.}~\bibnamefont
  {{Comelli}}}, \bibinfo {author} {\bibfnamefont {A.}~\bibnamefont {{Riotto}}},
  \bibinfo {author} {\bibfnamefont {F.}~\bibnamefont {{Sala}}}, \bibinfo
  {author} {\bibfnamefont {A.}~\bibnamefont {{Strumia}}},\ and\ \bibinfo
  {author} {\bibfnamefont {A.}~\bibnamefont {{Urbano}}},\ }\href
  {https://doi.org/10.1088/1475-7516/2011/03/019} {\bibfield  {journal}
  {\bibinfo  {journal} {{J. Cosmol. Astropart. Phys.}}\ }\textbf {\bibinfo
  {volume} {03}},\ \bibinfo {eid} {019} (\bibinfo {year} {2011})},\ \Eprint
  {https://arxiv.org/abs/1009.0224} {arXiv:1009.0224 [hep-ph]} \BibitemShut
  {NoStop}%
\bibitem [{\citenamefont {{Moskalenko}}\ and\ \citenamefont
  {{Strong}}(2000)}]{2000ApJ...528..357M}%
  \BibitemOpen
  \bibfield  {author} {\bibinfo {author} {\bibfnamefont {I.~V.}\ \bibnamefont
  {{Moskalenko}}}\ and\ \bibinfo {author} {\bibfnamefont {A.~W.}\ \bibnamefont
  {{Strong}}},\ }\href {https://doi.org/10.1086/308138} {\bibfield  {journal}
  {\bibinfo  {journal} {Astrophys. J.}\ }\textbf {\bibinfo {volume} {528}},\
  \bibinfo {pages} {357} (\bibinfo {year} {2000})},\ \Eprint
  {https://arxiv.org/abs/astro-ph/9811284} {arXiv:astro-ph/9811284 [astro-ph]}
  \BibitemShut {NoStop}%
\bibitem [{\citenamefont {{Porter}}\ and\ \citenamefont
  {{Strong}}(2005)}]{2005ICRC....4...77P}%
  \BibitemOpen
  \bibfield  {author} {\bibinfo {author} {\bibfnamefont {T.~A.}\ \bibnamefont
  {{Porter}}}\ and\ \bibinfo {author} {\bibfnamefont {A.~W.}\ \bibnamefont
  {{Strong}}},\ }in\ \href@noop {} {\emph {\bibinfo {booktitle} {29th
  International Cosmic Ray Conference (ICRC29), Volume 4}}},\ \bibinfo {series}
  {International Cosmic Ray Conference}, Vol.~\bibinfo {volume} {4}\ (\bibinfo
  {year} {2005})\ p.~\bibinfo {pages} {77}\BibitemShut {NoStop}%
\bibitem [{\citenamefont {{Yin}}\ \emph {et~al.}(2009)\citenamefont {{Yin}},
  \citenamefont {{Hou}}, \citenamefont {{Prantzos}}, \citenamefont
  {{Boissier}}, \citenamefont {{Chang}}, \citenamefont {{Shen}},\ and\
  \citenamefont {{Zhang}}}]{2009A&A...505..497Y}%
  \BibitemOpen
  \bibfield  {author} {\bibinfo {author} {\bibfnamefont {J.}~\bibnamefont
  {{Yin}}}, \bibinfo {author} {\bibfnamefont {J.~L.}\ \bibnamefont {{Hou}}},
  \bibinfo {author} {\bibfnamefont {N.}~\bibnamefont {{Prantzos}}}, \bibinfo
  {author} {\bibfnamefont {S.}~\bibnamefont {{Boissier}}}, \bibinfo {author}
  {\bibfnamefont {R.~X.}\ \bibnamefont {{Chang}}}, \bibinfo {author}
  {\bibfnamefont {S.~Y.}\ \bibnamefont {{Shen}}},\ and\ \bibinfo {author}
  {\bibfnamefont {B.}~\bibnamefont {{Zhang}}},\ }\href
  {https://doi.org/10.1051/0004-6361/200912316} {\bibfield  {journal} {\bibinfo
   {journal} {Astron. Astrophys.}\ }\textbf {\bibinfo {volume} {505}},\
  \bibinfo {pages} {497} (\bibinfo {year} {2009})},\ \Eprint
  {https://arxiv.org/abs/0906.4821} {arXiv:0906.4821 [astro-ph.GA]}
  \BibitemShut {NoStop}%
\bibitem [{\citenamefont {{Sodroski}}\ \emph {et~al.}(1997)\citenamefont
  {{Sodroski}}, \citenamefont {{Odegard}}, \citenamefont {{Arendt}},
  \citenamefont {{Dwek}}, \citenamefont {{Weiland}}, \citenamefont {{Hauser}},\
  and\ \citenamefont {{Kelsall}}}]{1997ApJ...480..173S}%
  \BibitemOpen
  \bibfield  {author} {\bibinfo {author} {\bibfnamefont {T.~J.}\ \bibnamefont
  {{Sodroski}}}, \bibinfo {author} {\bibfnamefont {N.}~\bibnamefont
  {{Odegard}}}, \bibinfo {author} {\bibfnamefont {R.~G.}\ \bibnamefont
  {{Arendt}}}, \bibinfo {author} {\bibfnamefont {E.}~\bibnamefont {{Dwek}}},
  \bibinfo {author} {\bibfnamefont {J.~L.}\ \bibnamefont {{Weiland}}}, \bibinfo
  {author} {\bibfnamefont {M.~G.}\ \bibnamefont {{Hauser}}},\ and\ \bibinfo
  {author} {\bibfnamefont {T.}~\bibnamefont {{Kelsall}}},\ }\href
  {https://doi.org/10.1086/303961} {\bibfield  {journal} {\bibinfo  {journal}
  {Astrophys. J.}\ }\textbf {\bibinfo {volume} {480}},\ \bibinfo {pages} {173}
  (\bibinfo {year} {1997})}\BibitemShut {NoStop}%
\bibitem [{\citenamefont {{Draine}}\ \emph {et~al.}(2014)\citenamefont
  {{Draine}}, \citenamefont {{Aniano}}, \citenamefont {{Krause}}, \citenamefont
  {{Groves}}, \citenamefont {{Sandstrom}}, \citenamefont {{Braun}},
  \citenamefont {{Leroy}}, \citenamefont {{Klaas}}, \citenamefont {{Linz}},
  \citenamefont {{Rix}}, \citenamefont {{Schinnerer}}, \citenamefont
  {{Schmiedeke}},\ and\ \citenamefont {{Walter}}}]{2014ApJ...780..172D}%
  \BibitemOpen
  \bibfield  {author} {\bibinfo {author} {\bibfnamefont {B.~T.}\ \bibnamefont
  {{Draine}}}, \bibinfo {author} {\bibfnamefont {G.}~\bibnamefont {{Aniano}}},
  \bibinfo {author} {\bibfnamefont {O.}~\bibnamefont {{Krause}}}, \bibinfo
  {author} {\bibfnamefont {B.}~\bibnamefont {{Groves}}}, \bibinfo {author}
  {\bibfnamefont {K.}~\bibnamefont {{Sandstrom}}}, \bibinfo {author}
  {\bibfnamefont {R.}~\bibnamefont {{Braun}}}, \bibinfo {author} {\bibfnamefont
  {A.}~\bibnamefont {{Leroy}}}, \bibinfo {author} {\bibfnamefont
  {U.}~\bibnamefont {{Klaas}}}, \bibinfo {author} {\bibfnamefont
  {H.}~\bibnamefont {{Linz}}}, \bibinfo {author} {\bibfnamefont {H.-W.}\
  \bibnamefont {{Rix}}}, \bibinfo {author} {\bibfnamefont {E.}~\bibnamefont
  {{Schinnerer}}}, \bibinfo {author} {\bibfnamefont {A.}~\bibnamefont
  {{Schmiedeke}}},\ and\ \bibinfo {author} {\bibfnamefont {F.}~\bibnamefont
  {{Walter}}},\ }\href {https://doi.org/10.1088/0004-637X/780/2/172} {\bibfield
   {journal} {\bibinfo  {journal} {Astrophys. J.}\ }\textbf {\bibinfo {volume}
  {780}},\ \bibinfo {eid} {172} (\bibinfo {year} {2014})},\ \Eprint
  {https://arxiv.org/abs/1306.2304} {arXiv:1306.2304 [astro-ph.CO]}
  \BibitemShut {NoStop}%
\bibitem [{\citenamefont {{Braun}}\ \emph {et~al.}(2009)\citenamefont
  {{Braun}}, \citenamefont {{Thilker}}, \citenamefont {{Walterbos}},\ and\
  \citenamefont {{Corbelli}}}]{2009ApJ...695..937B}%
  \BibitemOpen
  \bibfield  {author} {\bibinfo {author} {\bibfnamefont {R.}~\bibnamefont
  {{Braun}}}, \bibinfo {author} {\bibfnamefont {D.~A.}\ \bibnamefont
  {{Thilker}}}, \bibinfo {author} {\bibfnamefont {R.~A.~M.}\ \bibnamefont
  {{Walterbos}}},\ and\ \bibinfo {author} {\bibfnamefont {E.}~\bibnamefont
  {{Corbelli}}},\ }\href {https://doi.org/10.1088/0004-637X/695/2/937}
  {\bibfield  {journal} {\bibinfo  {journal} {Astrophys. J.}\ }\textbf
  {\bibinfo {volume} {695}},\ \bibinfo {pages} {937} (\bibinfo {year}
  {2009})},\ \Eprint {https://arxiv.org/abs/0901.4154} {arXiv:0901.4154
  [astro-ph.CO]} \BibitemShut {NoStop}%
\bibitem [{\citenamefont {{Nieten}}\ \emph {et~al.}(2006)\citenamefont
  {{Nieten}}, \citenamefont {{Neininger}}, \citenamefont {{Gu{\'e}lin}},
  \citenamefont {{Ungerechts}}, \citenamefont {{Lucas}}, \citenamefont
  {{Berkhuijsen}}, \citenamefont {{Beck}},\ and\ \citenamefont
  {{Wielebinski}}}]{2006A&A...453..459N}%
  \BibitemOpen
  \bibfield  {author} {\bibinfo {author} {\bibfnamefont {C.}~\bibnamefont
  {{Nieten}}}, \bibinfo {author} {\bibfnamefont {N.}~\bibnamefont
  {{Neininger}}}, \bibinfo {author} {\bibfnamefont {M.}~\bibnamefont
  {{Gu{\'e}lin}}}, \bibinfo {author} {\bibfnamefont {H.}~\bibnamefont
  {{Ungerechts}}}, \bibinfo {author} {\bibfnamefont {R.}~\bibnamefont
  {{Lucas}}}, \bibinfo {author} {\bibfnamefont {E.~M.}\ \bibnamefont
  {{Berkhuijsen}}}, \bibinfo {author} {\bibfnamefont {R.}~\bibnamefont
  {{Beck}}},\ and\ \bibinfo {author} {\bibfnamefont {R.}~\bibnamefont
  {{Wielebinski}}},\ }\href {https://doi.org/10.1051/0004-6361:20035672}
  {\bibfield  {journal} {\bibinfo  {journal} {Astron. Astrophys.}\ }\textbf
  {\bibinfo {volume} {453}},\ \bibinfo {pages} {459} (\bibinfo {year}
  {2006})},\ \Eprint {https://arxiv.org/abs/astro-ph/0512563}
  {arXiv:astro-ph/0512563 [astro-ph]} \BibitemShut {NoStop}%
\bibitem [{\citenamefont {{G{\'o}rski}}\ \emph {et~al.}(2005)\citenamefont
  {{G{\'o}rski}}, \citenamefont {{Hivon}}, \citenamefont {{Banday}},
  \citenamefont {{Wandelt}}, \citenamefont {{Hansen}}, \citenamefont
  {{Reinecke}},\ and\ \citenamefont {{Bartelmann}}}]{2005ApJ...622..759G}%
  \BibitemOpen
  \bibfield  {author} {\bibinfo {author} {\bibfnamefont {K.~M.}\ \bibnamefont
  {{G{\'o}rski}}}, \bibinfo {author} {\bibfnamefont {E.}~\bibnamefont
  {{Hivon}}}, \bibinfo {author} {\bibfnamefont {A.~J.}\ \bibnamefont
  {{Banday}}}, \bibinfo {author} {\bibfnamefont {B.~D.}\ \bibnamefont
  {{Wandelt}}}, \bibinfo {author} {\bibfnamefont {F.~K.}\ \bibnamefont
  {{Hansen}}}, \bibinfo {author} {\bibfnamefont {M.}~\bibnamefont
  {{Reinecke}}},\ and\ \bibinfo {author} {\bibfnamefont {M.}~\bibnamefont
  {{Bartelmann}}},\ }\href {https://doi.org/10.1086/427976} {\bibfield
  {journal} {\bibinfo  {journal} {Astrophys. J.}\ }\textbf {\bibinfo {volume}
  {622}},\ \bibinfo {pages} {759} (\bibinfo {year} {2005})},\ \Eprint
  {https://arxiv.org/abs/arXiv:astro-ph/0409513} {arXiv:astro-ph/0409513}
  \BibitemShut {NoStop}%
\bibitem [{Fer({\natexlab{a}})}]{Fermi}%
  \BibitemOpen
  \href@noop {} {}\bibinfo {howpublished}
  {\url{https://www.slac.stanford.edu/exp/glast/groups/canda/lat_Performance.htm}}
  ({\natexlab{a}})\BibitemShut {NoStop}%
\bibitem [{\citenamefont {{De Angelis}}\ \emph {et~al.}(2017)\citenamefont {{De
  Angelis}}, \citenamefont {{Tatischeff}}, \citenamefont {{Tavani}},
  \citenamefont {{Oberlack}}, \citenamefont {{Grenier}}, \citenamefont
  {{Hanlon}}, \citenamefont {{Walter}}, \citenamefont {{Argan}}, \citenamefont
  {{von Ballmoos}}, \citenamefont {{Bulgarelli}}, \citenamefont {{Donnarumma}},
  \citenamefont {{Hernanz}}, \citenamefont {{Kuvvetli}}, \citenamefont
  {{Pearce}}, \citenamefont {{Zdziarski}} \emph
  {et~al.}}]{2017ExA...tmp...24D}%
  \BibitemOpen
  \bibfield  {author} {\bibinfo {author} {\bibfnamefont {A.}~\bibnamefont {{De
  Angelis}}}, \bibinfo {author} {\bibfnamefont {V.}~\bibnamefont
  {{Tatischeff}}}, \bibinfo {author} {\bibfnamefont {M.}~\bibnamefont
  {{Tavani}}}, \bibinfo {author} {\bibfnamefont {U.}~\bibnamefont
  {{Oberlack}}}, \bibinfo {author} {\bibfnamefont {I.}~\bibnamefont
  {{Grenier}}}, \bibinfo {author} {\bibfnamefont {L.}~\bibnamefont {{Hanlon}}},
  \bibinfo {author} {\bibfnamefont {R.}~\bibnamefont {{Walter}}}, \bibinfo
  {author} {\bibfnamefont {A.}~\bibnamefont {{Argan}}}, \bibinfo {author}
  {\bibfnamefont {P.}~\bibnamefont {{von Ballmoos}}}, \bibinfo {author}
  {\bibfnamefont {A.}~\bibnamefont {{Bulgarelli}}}, \bibinfo {author}
  {\bibfnamefont {I.}~\bibnamefont {{Donnarumma}}}, \bibinfo {author}
  {\bibfnamefont {M.}~\bibnamefont {{Hernanz}}}, \bibinfo {author}
  {\bibfnamefont {I.}~\bibnamefont {{Kuvvetli}}}, \bibinfo {author}
  {\bibfnamefont {M.}~\bibnamefont {{Pearce}}}, \bibinfo {author}
  {\bibfnamefont {A.}~\bibnamefont {{Zdziarski}}}, \emph {et~al.},\ }\href
  {https://doi.org/10.1007/s10686-017-9533-6} {\bibfield  {journal} {\bibinfo
  {journal} {Exper. Astron.}\ }\textbf {\bibinfo {volume} {44}},\ \bibinfo
  {pages} {25} (\bibinfo {year} {2017})},\ \Eprint
  {https://arxiv.org/abs/1611.02232} {arXiv:1611.02232 [astro-ph.HE]}
  \BibitemShut {NoStop}%
\bibitem [{Fer({\natexlab{b}})}]{Fermi-map}%
  \BibitemOpen
  \href@noop {} {}\bibinfo {howpublished}
  {\url{https://fermi.gsfc.nasa.gov/ssc/data/access/lat/BackgroundModels.html}}
  ({\natexlab{b}})\BibitemShut {NoStop}%
\bibitem [{\citenamefont {{Tempel}}\ \emph {et~al.}(2010)\citenamefont
  {{Tempel}}, \citenamefont {{Tamm}},\ and\ \citenamefont
  {{Tenjes}}}]{2010A&A...509A..91T}%
  \BibitemOpen
  \bibfield  {author} {\bibinfo {author} {\bibfnamefont {E.}~\bibnamefont
  {{Tempel}}}, \bibinfo {author} {\bibfnamefont {A.}~\bibnamefont {{Tamm}}},\
  and\ \bibinfo {author} {\bibfnamefont {P.}~\bibnamefont {{Tenjes}}},\ }\href
  {https://doi.org/10.1051/0004-6361/200912186} {\bibfield  {journal} {\bibinfo
   {journal} {Astron. Astrophys.}\ }\textbf {\bibinfo {volume} {509}},\
  \bibinfo {eid} {A91} (\bibinfo {year} {2010})},\ \Eprint
  {https://arxiv.org/abs/0912.0124} {arXiv:0912.0124 [astro-ph.CO]}
  \BibitemShut {NoStop}%
\bibitem [{\citenamefont {{Li}}\ \emph {et~al.}(2021)\citenamefont {{Li}},
  \citenamefont {{Riess}}, \citenamefont {{Busch}}, \citenamefont
  {{Casertano}}, \citenamefont {{Macri}},\ and\ \citenamefont
  {{Yuan}}}]{2021ApJ...920...84L}%
  \BibitemOpen
  \bibfield  {author} {\bibinfo {author} {\bibfnamefont {S.}~\bibnamefont
  {{Li}}}, \bibinfo {author} {\bibfnamefont {A.~G.}\ \bibnamefont {{Riess}}},
  \bibinfo {author} {\bibfnamefont {M.~P.}\ \bibnamefont {{Busch}}}, \bibinfo
  {author} {\bibfnamefont {S.}~\bibnamefont {{Casertano}}}, \bibinfo {author}
  {\bibfnamefont {L.~M.}\ \bibnamefont {{Macri}}},\ and\ \bibinfo {author}
  {\bibfnamefont {W.}~\bibnamefont {{Yuan}}},\ }\href
  {https://doi.org/10.3847/1538-4357/ac1597} {\bibfield  {journal} {\bibinfo
  {journal} {Astrophys. J.}\ }\textbf {\bibinfo {volume} {920}},\ \bibinfo
  {eid} {84} (\bibinfo {year} {2021})},\ \Eprint
  {https://arxiv.org/abs/2107.08029} {arXiv:2107.08029 [astro-ph.CO]}
  \BibitemShut {NoStop}%
\bibitem [{\citenamefont {{McEnery}}\ \emph {et~al.}(2019)\citenamefont
  {{McEnery}}, \citenamefont {{van der Horst}}, \citenamefont {{Dominguez}},
  \citenamefont {{Moiseev}}, \citenamefont {{Marcowith}}, \citenamefont
  {{Harding}}, \citenamefont {{Lien}}, \citenamefont {{Giuliani}},
  \citenamefont {{Inglis}}, \citenamefont {{Ansoldi}}, \citenamefont
  {{Stamerra}}, \citenamefont {{Manousakis}}, \citenamefont {{Strong}},
  \citenamefont {{Bambi}}, \citenamefont {{Patricelli}} \emph
  {et~al.}}]{2019BAAS...51g.245M}%
  \BibitemOpen
  \bibfield  {author} {\bibinfo {author} {\bibfnamefont {J.}~\bibnamefont
  {{McEnery}}}, \bibinfo {author} {\bibfnamefont {A.}~\bibnamefont {{van der
  Horst}}}, \bibinfo {author} {\bibfnamefont {A.}~\bibnamefont {{Dominguez}}},
  \bibinfo {author} {\bibfnamefont {A.}~\bibnamefont {{Moiseev}}}, \bibinfo
  {author} {\bibfnamefont {A.}~\bibnamefont {{Marcowith}}}, \bibinfo {author}
  {\bibfnamefont {A.}~\bibnamefont {{Harding}}}, \bibinfo {author}
  {\bibfnamefont {A.}~\bibnamefont {{Lien}}}, \bibinfo {author} {\bibfnamefont
  {A.}~\bibnamefont {{Giuliani}}}, \bibinfo {author} {\bibfnamefont
  {A.}~\bibnamefont {{Inglis}}}, \bibinfo {author} {\bibfnamefont
  {S.}~\bibnamefont {{Ansoldi}}}, \bibinfo {author} {\bibfnamefont
  {A.}~\bibnamefont {{Stamerra}}}, \bibinfo {author} {\bibfnamefont
  {A.}~\bibnamefont {{Manousakis}}}, \bibinfo {author} {\bibfnamefont
  {A.}~\bibnamefont {{Strong}}}, \bibinfo {author} {\bibfnamefont
  {C.}~\bibnamefont {{Bambi}}}, \bibinfo {author} {\bibfnamefont
  {B.}~\bibnamefont {{Patricelli}}}, \emph {et~al.},\ }in\ \href
  {https://doi.org/10.48550/arXiv.1907.07558} {\emph {\bibinfo {booktitle}
  {Bull. Amer. Astron. Soc.}}},\ Vol.~\bibinfo {volume} {51}\ (\bibinfo {year}
  {2019})\ p.\ \bibinfo {pages} {245},\ \Eprint
  {https://arxiv.org/abs/1907.07558} {arXiv:1907.07558 [astro-ph.IM]}
  \BibitemShut {NoStop}%
\bibitem [{\citenamefont {{Fari{\~n}a}}\ \emph {et~al.}(2022)\citenamefont
  {{Fari{\~n}a}}, \citenamefont {{Jouvin}}, \citenamefont {{Rico}},
  \citenamefont {{Mori}}, \citenamefont {{Gargano}}, \citenamefont {{Formato}},
  \citenamefont {{de Palma}}, \citenamefont {{Pizzolotto}}, \citenamefont
  {{Casaus}}, \citenamefont {{Mazziota}}, \citenamefont {{Wu}}, \citenamefont
  {{Bordas}}, \citenamefont {{Gascon}}, \citenamefont {{Simons}}, \citenamefont
  {{Tykhonov}}, \citenamefont {{Altomare}}, \citenamefont {{Silveri}},
  \citenamefont {{Gasparrini}},\ and\ \citenamefont {{HERD
  Collaboration}}}]{2022icrc.confE.651F}%
  \BibitemOpen
  \bibfield  {author} {\bibinfo {author} {\bibfnamefont {L.}~\bibnamefont
  {{Fari{\~n}a}}}, \bibinfo {author} {\bibfnamefont {L.}~\bibnamefont
  {{Jouvin}}}, \bibinfo {author} {\bibfnamefont {J.}~\bibnamefont {{Rico}}},
  \bibinfo {author} {\bibfnamefont {N.}~\bibnamefont {{Mori}}}, \bibinfo
  {author} {\bibfnamefont {F.}~\bibnamefont {{Gargano}}}, \bibinfo {author}
  {\bibfnamefont {V.}~\bibnamefont {{Formato}}}, \bibinfo {author}
  {\bibfnamefont {F.}~\bibnamefont {{de Palma}}}, \bibinfo {author}
  {\bibfnamefont {C.}~\bibnamefont {{Pizzolotto}}}, \bibinfo {author}
  {\bibfnamefont {J.}~\bibnamefont {{Casaus}}}, \bibinfo {author}
  {\bibfnamefont {M.~N.}\ \bibnamefont {{Mazziota}}}, \bibinfo {author}
  {\bibfnamefont {X.}~\bibnamefont {{Wu}}}, \bibinfo {author} {\bibfnamefont
  {P.}~\bibnamefont {{Bordas}}}, \bibinfo {author} {\bibfnamefont
  {D.}~\bibnamefont {{Gascon}}}, \bibinfo {author} {\bibfnamefont
  {D.}~\bibnamefont {{Simons}}}, \bibinfo {author} {\bibfnamefont
  {A.}~\bibnamefont {{Tykhonov}}}, \bibinfo {author} {\bibfnamefont
  {C.}~\bibnamefont {{Altomare}}}, \bibinfo {author} {\bibfnamefont
  {L.}~\bibnamefont {{Silveri}}}, \bibinfo {author} {\bibfnamefont
  {D.}~\bibnamefont {{Gasparrini}}},\ and\ \bibinfo {author} {\bibnamefont
  {{HERD Collaboration}}},\ }in\ \href@noop {} {\emph {\bibinfo {booktitle}
  {37th International Cosmic Ray Conference. 12-23 July 2021. Berlin}}}\
  (\bibinfo {year} {2022})\ p.\ \bibinfo {pages} {651}\BibitemShut {NoStop}%
\bibitem [{\citenamefont {{Egorov}}\ \emph {et~al.}(2018)\citenamefont
  {{Egorov}}, \citenamefont {{Galper}}, \citenamefont {{Topchiev}},
  \citenamefont {{Leonov}}, \citenamefont {{Suchkov}}, \citenamefont
  {{Kheymits}},\ and\ \citenamefont {{Yurkin}}}]{2018PAN....81..373E}%
  \BibitemOpen
  \bibfield  {author} {\bibinfo {author} {\bibfnamefont {A.~E.}\ \bibnamefont
  {{Egorov}}}, \bibinfo {author} {\bibfnamefont {A.~M.}\ \bibnamefont
  {{Galper}}}, \bibinfo {author} {\bibfnamefont {N.~P.}\ \bibnamefont
  {{Topchiev}}}, \bibinfo {author} {\bibfnamefont {A.~A.}\ \bibnamefont
  {{Leonov}}}, \bibinfo {author} {\bibfnamefont {S.~I.}\ \bibnamefont
  {{Suchkov}}}, \bibinfo {author} {\bibfnamefont {M.~D.}\ \bibnamefont
  {{Kheymits}}},\ and\ \bibinfo {author} {\bibfnamefont {Y.~T.}\ \bibnamefont
  {{Yurkin}}},\ }\href {https://doi.org/10.1134/S1063778818030110} {\bibfield
  {journal} {\bibinfo  {journal} {Phys. Atom. Nucl.}\ }\textbf {\bibinfo
  {volume} {81}},\ \bibinfo {pages} {373} (\bibinfo {year} {2018})},\ \Eprint
  {https://arxiv.org/abs/1710.02492} {arXiv:1710.02492 [astro-ph.HE]}
  \BibitemShut {NoStop}%
\bibitem [{\citenamefont {{Egorov}}\ \emph {et~al.}(2020)\citenamefont
  {{Egorov}}, \citenamefont {{Topchiev}}, \citenamefont {{Galper}},
  \citenamefont {{Dalkarov}}, \citenamefont {{Leonov}}, \citenamefont
  {{Suchkov}},\ and\ \citenamefont {{Yurkin}}}]{2020JCAP...11..049E}%
  \BibitemOpen
  \bibfield  {author} {\bibinfo {author} {\bibfnamefont {A.~E.}\ \bibnamefont
  {{Egorov}}}, \bibinfo {author} {\bibfnamefont {N.~P.}\ \bibnamefont
  {{Topchiev}}}, \bibinfo {author} {\bibfnamefont {A.~M.}\ \bibnamefont
  {{Galper}}}, \bibinfo {author} {\bibfnamefont {O.~D.}\ \bibnamefont
  {{Dalkarov}}}, \bibinfo {author} {\bibfnamefont {A.~A.}\ \bibnamefont
  {{Leonov}}}, \bibinfo {author} {\bibfnamefont {S.~I.}\ \bibnamefont
  {{Suchkov}}},\ and\ \bibinfo {author} {\bibfnamefont {Y.~T.}\ \bibnamefont
  {{Yurkin}}},\ }\href {https://doi.org/10.1088/1475-7516/2020/11/049}
  {\bibfield  {journal} {\bibinfo  {journal} {{J. Cosmol. Astropart. Phys.}}\
  }\textbf {\bibinfo {volume} {11}},\ \bibinfo {eid} {049} (\bibinfo {year}
  {2020})},\ \Eprint {https://arxiv.org/abs/2005.09032} {arXiv:2005.09032
  [astro-ph.HE]} \BibitemShut {NoStop}%
\bibitem [{\citenamefont {{Djuvsland}}\ \emph {et~al.}(2023)\citenamefont
  {{Djuvsland}}, \citenamefont {{Hinton}},\ and\ \citenamefont
  {{Reville}}}]{2023PDU....3901157D}%
  \BibitemOpen
  \bibfield  {author} {\bibinfo {author} {\bibfnamefont {J.~I.}\ \bibnamefont
  {{Djuvsland}}}, \bibinfo {author} {\bibfnamefont {J.}~\bibnamefont
  {{Hinton}}},\ and\ \bibinfo {author} {\bibfnamefont {B.}~\bibnamefont
  {{Reville}}},\ }\href {https://doi.org/10.1016/j.dark.2022.101157} {\bibfield
   {journal} {\bibinfo  {journal} {Phys. Dark Univ.}\ }\textbf {\bibinfo
  {volume} {39}},\ \bibinfo {eid} {101157} (\bibinfo {year} {2023})},\ \Eprint
  {https://arxiv.org/abs/2212.05785} {arXiv:2212.05785 [astro-ph.HE]}
  \BibitemShut {NoStop}%
\bibitem [{\citenamefont {{Boldrini}}\ \emph {et~al.}(2021)\citenamefont
  {{Boldrini}}, \citenamefont {{Mohayaee}},\ and\ \citenamefont
  {{Silk}}}]{2021ApJ...919...86B}%
  \BibitemOpen
  \bibfield  {author} {\bibinfo {author} {\bibfnamefont {P.}~\bibnamefont
  {{Boldrini}}}, \bibinfo {author} {\bibfnamefont {R.}~\bibnamefont
  {{Mohayaee}}},\ and\ \bibinfo {author} {\bibfnamefont {J.}~\bibnamefont
  {{Silk}}},\ }\href {https://doi.org/10.3847/1538-4357/ac12d3} {\bibfield
  {journal} {\bibinfo  {journal} {Astrophys. J.}\ }\textbf {\bibinfo {volume}
  {919}},\ \bibinfo {eid} {86} (\bibinfo {year} {2021})},\ \Eprint
  {https://arxiv.org/abs/2002.12192} {arXiv:2002.12192 [astro-ph.GA]}
  \BibitemShut {NoStop}%
\bibitem [{\citenamefont {{Saikawa}}\ and\ \citenamefont
  {{Shirai}}(2020)}]{2020JCAP...08..011S}%
  \BibitemOpen
  \bibfield  {author} {\bibinfo {author} {\bibfnamefont {K.}~\bibnamefont
  {{Saikawa}}}\ and\ \bibinfo {author} {\bibfnamefont {S.}~\bibnamefont
  {{Shirai}}},\ }\href {https://doi.org/10.1088/1475-7516/2020/08/011}
  {\bibfield  {journal} {\bibinfo  {journal} {{J. Cosmol. Astropart. Phys.}}\
  }\textbf {\bibinfo {volume} {08}},\ \bibinfo {eid} {011} (\bibinfo {year}
  {2020})},\ \Eprint {https://arxiv.org/abs/2005.03544} {arXiv:2005.03544
  [hep-ph]} \BibitemShut {NoStop}%
\bibitem [{\citenamefont {{Ghosh}}\ \emph {et~al.}(2021)\citenamefont
  {{Ghosh}}, \citenamefont {{Dutta Banik}}, \citenamefont {{Chun}},\ and\
  \citenamefont {{Majumdar}}}]{2021PhRvD.104g5016G}%
  \BibitemOpen
  \bibfield  {author} {\bibinfo {author} {\bibfnamefont {S.}~\bibnamefont
  {{Ghosh}}}, \bibinfo {author} {\bibfnamefont {A.}~\bibnamefont {{Dutta
  Banik}}}, \bibinfo {author} {\bibfnamefont {E.~J.}\ \bibnamefont {{Chun}}},\
  and\ \bibinfo {author} {\bibfnamefont {D.}~\bibnamefont {{Majumdar}}},\
  }\href {https://doi.org/10.1103/PhysRevD.104.075016} {\bibfield  {journal}
  {\bibinfo  {journal} {Phys. Rev. D}\ }\textbf {\bibinfo {volume} {104}},\
  \bibinfo {eid} {075016} (\bibinfo {year} {2021})},\ \Eprint
  {https://arxiv.org/abs/2003.07675} {arXiv:2003.07675 [hep-ph]} \BibitemShut
  {NoStop}%
\bibitem [{\citenamefont {{Bonnarel}}\ \emph {et~al.}(2000)\citenamefont
  {{Bonnarel}}, \citenamefont {{Fernique}}, \citenamefont {{Bienaym{\'e}}},
  \citenamefont {{Egret}}, \citenamefont {{Genova}}, \citenamefont {{Louys}},
  \citenamefont {{Ochsenbein}}, \citenamefont {{Wenger}},\ and\ \citenamefont
  {{Bartlett}}}]{2000A&AS..143...33B}%
  \BibitemOpen
  \bibfield  {author} {\bibinfo {author} {\bibfnamefont {F.}~\bibnamefont
  {{Bonnarel}}}, \bibinfo {author} {\bibfnamefont {P.}~\bibnamefont
  {{Fernique}}}, \bibinfo {author} {\bibfnamefont {O.}~\bibnamefont
  {{Bienaym{\'e}}}}, \bibinfo {author} {\bibfnamefont {D.}~\bibnamefont
  {{Egret}}}, \bibinfo {author} {\bibfnamefont {F.}~\bibnamefont {{Genova}}},
  \bibinfo {author} {\bibfnamefont {M.}~\bibnamefont {{Louys}}}, \bibinfo
  {author} {\bibfnamefont {F.}~\bibnamefont {{Ochsenbein}}}, \bibinfo {author}
  {\bibfnamefont {M.}~\bibnamefont {{Wenger}}},\ and\ \bibinfo {author}
  {\bibfnamefont {J.~G.}\ \bibnamefont {{Bartlett}}},\ }\href
  {https://doi.org/10.1051/aas:2000331} {\bibfield  {journal} {\bibinfo
  {journal} {Astron. Astrophys. Suppl. Ser.}\ }\textbf {\bibinfo {volume}
  {143}},\ \bibinfo {pages} {33} (\bibinfo {year} {2000})}\BibitemShut
  {NoStop}%
\bibitem [{num()}]{numpy}%
  \BibitemOpen
  \href@noop {} {}\bibinfo {howpublished}
  {\url{https://numpy.org/}}\BibitemShut {NoStop}%
\bibitem [{hea()}]{healpy}%
  \BibitemOpen
  \href@noop {} {}\bibinfo {howpublished}
  {\url{https://github.com/healpy}}\BibitemShut {NoStop}%
\bibitem [{WPD()}]{WPD}%
  \BibitemOpen
  \href@noop {} {}\bibinfo {howpublished}
  {\url{https://apps.automeris.io/wpd/}}\BibitemShut {NoStop}%
\end{thebibliography}%

\end{document}